\documentclass[twocolumn,floatfix, nofootinbib,prd,aps,preprintnumbers,showpacs,showkeys,superscriptaddress,longbibliography]{revtex4-1}

\usepackage[mathscr]{euscript}
\usepackage{amsmath}
\usepackage{graphicx}
\usepackage{dcolumn}
\usepackage{bm}
\usepackage{epsfig}
\usepackage{amssymb,latexsym,mathrsfs}
\usepackage{graphicx}
\usepackage{color}
\usepackage{hyperref}
\usepackage{float}
\usepackage{diagbox}

\usepackage{tikz}

\usepackage{amssymb}
\usepackage{pifont}%

\usepackage{amsmath}
\usepackage{physics}
\usepackage[style=english]{csquotes} 

\usepackage{lipsum, babel}
\usepackage{amssymb}
\usepackage{color}
\usepackage{hyperref}
\usepackage{bm}
\usepackage{amsfonts}
\usepackage{mathrsfs}
\usepackage{graphicx}
\usepackage{amsmath}
\usepackage{float}
\usepackage{braket}
\usepackage{natbib}

     \hypersetup{
    colorlinks=true,
    linkcolor=red,
    citecolor=blue,
} 

\usepackage[caption=false]{subfig}

\usepackage{amssymb}
\usepackage[mathscr]{euscript}
\usepackage{amsmath}
\usepackage{graphicx}
\usepackage{dcolumn}
\usepackage{bm}
\usepackage{epsfig}
\usepackage{amssymb,latexsym,mathrsfs}
\usepackage{graphicx}
\usepackage{color}
\usepackage{hyperref}
\usepackage{float}
\usepackage[thinlines]{easytable}
\usepackage{multirow}
\usepackage{diagbox}
\usepackage{varwidth}
\usepackage{array}
\newcolumntype{P}[1]{>{\centering\arraybackslash}p{#1}}

\newcolumntype{M}[1]{>{\centering\arraybackslash}m{#1}}
\newcolumntype{N}{@{}m{0pt}@{}}

\usepackage{amsmath}

\usepackage{amssymb}
\usepackage{pifont}
\hypersetup{
    colorlinks=true,
    linkcolor=red,
    citecolor=blue,
}

\usepackage{amsmath}
\usepackage{amssymb}
\usepackage[caption=false]{subfig}
\usepackage{hyperref}
\usepackage{url}
\usepackage{xcolor}
\usepackage{color}
\definecolor{amaranth}{rgb}{0.9, 0.17, 0.31}
\definecolor{purple(munsell)}{rgb}{0.62, 0.0, 0.77}
\definecolor{americanrose}{rgb}{1.0, 0.01, 0.24}
\definecolor{palatinateblue}{rgb}{0.15, 0.23, 0.89}
\definecolor{royalblue(web)}{rgb}{0.25, 0.41, 0.88}
\definecolor{hanpurple}{rgb}{0.32, 0.09, 0.98}
\definecolor{beaublue}{rgb}{0.74, 0.83, 0.9}
\definecolor{carminered}{rgb}{1.0, 0.0, 0.22}
\definecolor{brightpink}{rgb}{1.0, 0.0, 0.5}
\definecolor{vividviolet}{rgb}{0.62, 0.0, 1.0}
\definecolor{Uniblue}{RGB}{0,4,100}
\definecolor{crimson}{RGB}{180,0,20}

\definecolor{electron}{rgb}{1.0, 0.67, 0.22}
\hypersetup{ linktoc=all,
    colorlinks, linkcolor={Uniblue},
    citecolor={brightpink}, urlcolor={brightpink}}

\usepackage{physics}
\usepackage{csquotes}
\usepackage{mathtools}
\usepackage{soul}
\makeatletter
\def\l@subsubsection#1#2{}
\makeatother

\begin{document}

\newcommand{\sgn}{\operatorname{sgn}}
\newcommand{\hhat}[1]{\hat {\hat{#1}}}
\newcommand{\pslash}[1]{#1\llap{\sl/}}
\newcommand{\kslash}[1]{\rlap{\sl/}#1}
\newcommand{\lab}[1]{}
\newcommand{\sto}[1]{\begin{center} \textit{#1} \end{center}}
\newcommand{\rf}[1]{{\color{blue}[\textit{#1}]}}
\newcommand{\eml}[1]{#1}
\newcommand{\el}[1]{\label{#1}}
\newcommand{\er}[1]{Eq.\eqref{#1}}
\newcommand{\df}[1]{\textbf{#1}}
\newcommand{\mdf}[1]{\pmb{#1}}
\newcommand{\ft}[1]{\footnote{#1}}
\newcommand{\n}[1]{$#1$}
\newcommand{\fals}[1]{$^\times$ #1}
\newcommand{\new}{{\color{red}$^{NEW}$ }}
\newcommand{\ci}[1]{}
\newcommand{\de}[1]{{\color{green}\underline{#1}}}
\newcommand{\ke}{\rangle}
\newcommand{\br}{\langle}
\newcommand{\lb}{\left(}
\newcommand{\rb}{\right)}
\newcommand{\lbk}{\left[}
\newcommand{\rbk}{\right]}
\newcommand{\blb}{\Big(}
\newcommand{\brb}{\Big)}
\newcommand{\nn}{\nonumber \\}
\newcommand{\p}{\partial}
\newcommand{\pd}[1]{\frac {\partial} {\partial #1}}
\newcommand{\cd}{\nabla}
\newcommand{\cc}{$>$}
\newcommand{\bqa}{\begin{eqnarray}}
\newcommand{\eqa}{\end{eqnarray}}
\newcommand{\bqe}{\begin{equation}}
\newcommand{\eqe}{\end{equation}}
\newcommand{\bay}[1]{\left(\begin{array}{#1}}
\newcommand{\eay}{\end{array}\right)}
\newcommand{\eg}{\textit{e.g.} }
\newcommand{\ie}{\textit{i.e.}, }
\newcommand{\iv}[1]{{#1}^{-1}}

\newcommand{\at}[1]{{\Big|}_{#1}}
\newcommand{\zt}[1]{\texttt{#1}}
\newcommand{\non}{\nonumber}
\newcommand{\m}{\mu}
\def\xa{{m}}
\def\xA{{m}}
\def\xb{{\beta}}
\def\xB{{\Beta}}
\def\xd{{\delta}}
\def\xD{{\Delta}}
\def\xe{{\epsilon}}
\def\xE{{\Epsilon}}
\def\xve{{\varepsilon}}
\def\xg{{\gamma}}
\def\xG{{\Gamma}}
\def\xk{{\kappa}}
\def\xK{{\Kappa}}
\def\xl{{\lambda}}
\def\xL{{\Lambda}}
\def\xo{{\omega}}
\def\xO{{\Omega}}
\def\xvp{{\varphi}}
\def\xs{{\sigma}}
\def\xS{{\Sigma}}
\def\xt{{\theta}}
\def\xvt{{\vartheta}}
\def\xT{{\Theta}}
\def \Tr {{\rm Tr}}
\def\CA{{\cal A}}
\def\CC{{\cal C}}
\def\CD{{\cal D}}
\def\CE{{\cal E}}
\def\CF{{\cal F}}
\def\CH{{\cal H}}
\def\CJ{{\cal J}}
\def\CK{{\cal K}}
\def\CL{{\cal L}}
\def\CM{{\cal M}}
\def\CN{{\cal N}}
\def\CO{{\cal O}}
\def\CP{{\cal P}}
\def\CQ{{\cal Q}}
\def\CR{{\cal R}}
\def\CS{{\cal S}}
\def\CT{{\cal T}}
\def\CV{{\cal V}}
\def\CW{{\cal W}}
\def\CY{{\cal Y}}
\def\BC{\mathbb{C}}
\def\BR{\mathbb{R}}
\def\BZ{\mathbb{Z}}
\def\sA{\mathscr{A}}
\def\sB{\mathscr{B}}
\def\sF{\mathscr{F}}
\def\sG{\mathscr{G}}
\def\sH{\mathscr{H}}
\def\sJ{\mathscr{J}}
\def\sL{\mathscr{L}}
\def\sM{\mathscr{M}}
\def\sN{\mathscr{N}}
\def\sO{\mathscr{O}}
\def\sP{\mathscr{P}}
\def\sR{\mathscr{R}}
\def\sQ{\mathscr{Q}}
\def\sS{\mathscr{S}}
\def\sX{\mathscr{X}}

\def\slz{SL(2,Z)}
\def\slr{$SL(2,R)\times SL(2,R)$ }
\def\ads{${AdS}_5\times {S}^5$ }
\def\adst{${AdS}_3$ }
\def\sun{SU(N)}
\def\ad#1#2{{\frac \delta {\delta\sigma^{#1}} (#2)}}
\def\bqf{\bar Q_{\bar f}}
\def\nf{N_f}
\def\sunf{SU(N_f)}

\def\dcirc{{^\circ_\circ}}

\author{Morgan H. Lynch}
\email{morganlynch1984@gmail.com}
\affiliation{Center for Theoretical Physics,
Seoul National University, Seoul 08826, Korea}

\title{Experimental observation of a Rindler horizon}
\date{\today}

\begin{abstract}
In this manuscript we confirm the presence of a Rindler horizon at CERN-NA63 by exploring its thermodynamics induced by the Unruh effect in their high energy channeling radiation experiments. By linking the entropy of the emitted radiation to the photon number, we find the measured spectrum to be a simple manifestation of the second law of Rindler horizon thermodynamics and thus a direct measurement of the recoil Fulling-Davies-Unruh (FDU) temperature. Moreover, since the experiment is born out of an ultra-relativistic positron, and the FDU temperature is defined in the proper frame, we find that temperature boosts as a length and thus fast objects appear colder. The spectrum also provides us with a simple setting to measure fundamental constants, and we employ it to measure the positron mass. 
\end{abstract}


\maketitle

\section{Introduction}

The pursuits of quantum field theory in curved spacetime have led to a wide array of surprising discoveries. Most notably, it completely changed our understanding of such simple questions as how many particles are present in a given system. This ambiguity in particle number gave rise to a new class of kinematic particle production induced by horizons; namely the Parker, Hawking, and the Fulling-Davies-Unruh effects \cite{Parker:1968mv, hawking1974black, Davies:1976hi, Davies:1977yv, Unruh:1976db,lynch2021experimental, 2024PTEP.2024b3D01L}. Most surprising of all is the fact that the particles which are produced via these effects are thermalized at a temperature which is characterized by the acceleration scale of the system. In all cases, this acceleration characterizes the location of a horizon, either real or apparent, which is also present. 

When considering these horizons, one can also ascribe thermodynamic quantities to the system. Most notably, via the second law of thermodynamics, the connection between the area of the horizon, or area change, and the entropy encoded by it \cite{Bekenstein:1973ur, hawking1974black, bianchi2013mechanical}. All of these notions combine to paint a rather surprising picture. For an accelerated observer, the presence of these thermalized particles, along with the energy/momentum flux of particles through the associated horizon, gives rise to an area change in the horizon, which is completely determined by general relativity, i.e. the Einstein equation is the thermodynamic equation of state of these effervescent particles \cite{1995PhRvL..75.1260J}. The implication being that gravitation is an emergent quantity born out of these vacuum fluctuations. 

The recent experimental observation of radiation reaction and the Unruh effect by CERN-NA63 \cite{2018NatCo...9..795W, lynch2021experimental} provides us with an unprecedented opportunity to investigate the various tenets of quantum field theory in curved spacetime, e.g. a Hawking analysis of a black hole Rindler bath \cite{2024PhRvD.109j5009L}. In particular, we will now examine the thermodynamics of the Rindler horizon present in these systems. In the NA63 experiments \cite{2018NatCo...9..795W, 2019PhRvR...1c3014W}, a beam of high energy positrons, $E \sim 100$ GeV, penetrate samples of single crystal silicon. These positrons propagate along a crystalline axis of symmetry and are confined by a transverse potential created by the positively charged atomic sites. An immense Lorentz factor, $\gamma \sim 10^{5}$, enhances the local electric fields of the atomic sites and brings the local field strengths into, and beyond, the strong field regime \cite{2012RvMP...84.1177D}. The resultant radiation yields photons with energies, $\omega$, on the order of the initial positron beam energy, $\omega \sim E$. Then, from relativistic Newtons law, the proper recoil acceleration is subsequently set by the recoil kinetic energy of the photon emission, $a\sim \frac{ \hbar \omega^2}{mc}$. This then renders the Unruh effect, and Fulling-Davies-Unruh (FDU) temperature, $T_{FDU} = \frac{a \hbar}{2 \pi c k_{B}}$, accessible experimentally \cite{2017PhRvL.118p1102C, lynch2021experimental, 2025arXiv250521292L}. Here, we shall examine the resultant Rindler horizon thermodynamics which manifests in a suite of 7 NA63 data sets of silicon crystals with varying thickness and beam energy \cite{2018NatCo...9..795W, 2019PhRvR...1c3014W}. 

\section{The Unruh Threshold $\&$ Critical Frequency}
The channeling radiation experiments at NA63 function as high energy synchrotron emitters which probe the regime of quantum radiation reaction \cite{2018NatCo...9..795W}. The phenomenon of quantum radiation reaction is best understood as recoil produced by a hard photon with energy on the order of the rest mass of the positron. This recoil acceleration, is directly responsible for the unset of the Unruh effect. In order to examine the presence of the Unruh effect, and associated Rindler horizon, we must understand at what regime the Unruh effect saturates the dynamics. Let us examine the asymptotic time scales associated with synchrotron emission both with and without a hyperbolic/Unruh recoil trajectory \cite{2025arXiv250521292L}.

For radiative processes, the asymptotic power spectra can be computed via the stationary phase approximation \cite{1986SvPhU..29..788A}. Thus, given the amplitude, $\mathcal{A}$, of a radiative process,
\bqe
\mathcal{A} \sim \int dt e^{i(\textbf{k} \cdot \textbf{x} - \omega t)/\hbar},
\eqe
we will employ the stationary phase approximation. Here, we have a photon momentum and frequency, $\textbf{k}$ and $\omega$, emitted by a positron with a trajectory characterized by, $\textbf{x} = \textbf{x}(t) $. The condition for a stationary photon phase, $\phi =(\textbf{k}\cdot\textbf{x} - \omega t)/\hbar$, is given by $\frac{d\phi}{dt} = 0$. From this we have, $1-\beta(t) \cos{(\theta)}$. For radiation emission from relativistic particles, we have an emission angle, $\sin{(\theta)} \sim \frac{1}{\gamma}$, and we can thereby set, $\cos{(\theta)} \sim \beta(t)$, to arrive at our stationary phase solution. Hence,
\bqa
1-\beta^2(t_{0})  = 0 \non \\
\Rightarrow t_{0} = t_{1} + i t_{f}.  \label{phase}
\eqa
Here, the velocity, $\beta(t)$, characterize the semiclassical trajectory of the positron. In the above complex solution for the time, $t_{0} = t_{1} + i t_{f}$, $t_{1}$ determines the time at which the radiation is emitted and the formation time, $t_{f}$, determines the interval over which the radiation is emitted in phase. The coherence time, $t_{c}$, is determined by $t_1$ and yields the total timescale when the radiation is emitted coherently; 
\bqe
t_{c} = \frac{1}{\omega(1-v^2(t_{1}))}.  \label{tform}
\eqe
The subsequent radiation amplitude is then given by, $\mathcal{A} \sim e^{-t_{f}/t_{c}}$. The square of this amplitude then yields the overall power spectrum of the emitted radiation. Hence,
\bqe
\frac{d\mathcal{P}}{d\omega} \sim e^{-2t_{f}/t_{c}}. \label{pow}
\eqe
Let us now examine the cases for synchrotron emission both with and without a hyperbolic recoil acceleration. For the case of synchrotron emission, we have a trajectory characterized by $v_{z} = v_{0}\cos{(\Omega t)}$ and $v_{\perp} = -v_{0}\sin{(\Omega t)}$. Here, $\Omega$, is the channeling/synchrotron frequency. The resultant formation and coherence times are then given by \cite{1986SvPhU..29..788A},
\bqa
t_{f\Omega} &=& \frac{1}{\Omega \gamma} \non \\
t_{c \Omega} &=&  \frac{\gamma^2}{\omega}.
\eqa
Here the Lorentz gamma is determined by the beam velocity, $\gamma = 1/\sqrt{1-v_{0}^2}$. For the case of synchrotron emission with a hyperbolic/Unruh recoil we will make use of relativistic velocity addition. With the recoil trajectory, as measured in the beam frame given by, $v_{a} =  \frac{at/\gamma}{\sqrt{1 + \lb \frac{at}{\gamma } \rb^{2}}}$, we will then have a lab frame velocity given by $v=\frac{v_{z}-v_{a}}{1-v_{z}v_{a}}$. As such, for the Unruh case, the formation and coherence times are given by \cite{2025arXiv250521292L},
\bqa
t_{fa} &=& \frac{\gamma}{2a} \non \\
t_{c a} &=& \frac{7}{4\omega}.
\eqa
By comparing the resultant asymptotic power spectra, Eqn. (\ref{pow}), we require $\frac{d\mathcal{P}_{a}}{d\omega}  >  \frac{d\mathcal{P}_{\Omega}}{d\omega}$ for the Unruh effect to dominate. Utilizing the above formation and coherence times, this leads to the condition $\frac{2\gamma \omega}{7a} < \frac{\omega}{\Omega \gamma^3}$. Finally, by recalling $a = \frac{ \hbar \omega^2}{mc}$ and mapping the channeling oscillation to the characteristic photon frequency, $\omega = 2 \Omega \gamma^3$ \cite{2012RvMP...84.1177D}, we arrive at the condition where the hyperbolic recoil/Unruh effect dominates. Hence,
\bqe
\hbar \omega_c > \frac{E}{7}. \label{cond}
\eqe 
Here, $E$, is the initial beam energy of the positrons. Thus, for photons emitted above this critical frequency, the Unruh effect produced by the hyperbolic recoil will dominate the spectrum. Finally, in consideration of the suite of 7 NA63 channeling radiation experiments \cite{2019PhRvR...1c3014W, 2018NatCo...9..795W}, we will have the following two positron beam energies: 50 and 178.2 GeV. As such, our critical frequencies, above which the Unruh effect saturates the spectrum, will be given by \cite{2025arXiv250521292L},
\bqa
\hbar \omega_{178.2} &=& 25.5 \; \text{GeV} \non \\
\hbar \omega_{50} &=& 7.1 \; \text{GeV}.
\eqa
Let us now turn to a discussion of Rindler horizon thermodynamics and how it can be mapped to the emitted photon spectrum.

\section{Rindler horizon thermodynamics}

Thermodynamics offers a surprisingly simple technique to examine the properties of both black holes as well as Rindler horizons \cite{Bekenstein:1973ur, 1995PhRvL..75.1260J, 2013PhRvD..87l4031B}. In the high energy channeling radiation reaction experiments of CERN-NA63, Rindler horizons are rendered experimentally accessible via radiation reaction. There, we have a FDU temperature which is set by the immense recoil acceleration, $T_{FDU} = \frac{(\hbar \omega)^2}{2 \pi mc^2k_{B}}$. Systems that are thermalized at this temperature will then, by necessity, obey the second law of thermodynamics,
\bqe
dQ = k_{B}T dS. \label{2nd}
\eqe
Here, $dQ$, is the differential heat/energy flux which crosses the horizon, $dS$, is the resultant entropy change created by that flux. As we shall see, this statement of the second law not only describes the relationship between energy flux and entropy change, but it also describes the spectrum of thermalized particles due to the Unruh effect. Ultimately, this second law, and its application to Rindler horizons, lies at the heart of our understanding of gravity. As it has been shown \cite{1995PhRvL..75.1260J}, this second law along with the Bekenstein-Hawking area-entropy relation, $S = \frac{A}{4 \ell_{P}^2}$, with $\ell_{P}^{2}$ being the Planck area, is equivalent to the Einstein equation of general relativity.

\subsection{The Rindler energy}

The second law of thermodynamics, Eqn. (\ref{2nd}), for the case of total heat flux and entropy generation, can be when mapped to the case of the Rindler horizon via $\Delta Q = \Delta E$ and $\Delta S = \frac{\Delta A}{4 \ell_{P}^2}$. Utilizing $T_{FDU} = \frac{a \hbar }{2 \pi c k_{B}}$, the resulting expression of the second law is then given by \cite{2013PhRvD..87l4031B},
\bqe
\Delta A = \frac{8 \pi G \Delta E}{a c^2}.
\eqe
We must first comment on the fact that the area change for the Rindler horizon is equivalent to the area change for the event horizon of a black hole; subject to the condition that the energy deposited into the horizons and the acceleration/surface gravity are equivalent. Moreover, despite the fact that the total area of a Rindler horizon is infinite, the area change is indeed well defined.

In the Rindler setting, the energy, $\Delta E$, which fluxes through the Rindler horizon is defined by the Rindler observer, i.e. the energy measured in the accelerated reference frame. When examining the transition rate of an Unruh-DeWitt detector, this also plays the role of the energy gap \cite{2017PhRvL.118p1102C}. In the case of a synchrotron trajectory with a hyperbolic recoil, this energy gap/heat flux will be fixed by the transverse velocity, $Q = \beta_{\perp} \hbar \omega$, as measured in the Rindler frame \cite{2020PhRvD.101f5012P}. As such, the differential heat flux is given by  $dQ = \beta_{\perp} d(\hbar \omega)$. Our transverse velocity can thus be determined from the synchrotron trajectory, $\beta_{\perp} = \beta_{0} \gamma \sin{(\Omega t)}$. Note, we have included an additional factor of Lorentz gamma to boost into the beam frame. Here we shall consider the critical channeling oscillation frequency, $\Omega_c = \frac{\omega_c}{2 \gamma^3}$. For the oscillation time, we will have an upper bound set by the channeling formation time, as such we shall take the minimum $t = \frac{\gamma^2}{2E}$. Here the factor of 2 is for later convenience. Thus, we have $ \beta_{0}\sin{(\Omega_c t}) = \frac{1}{28 \gamma}$. Note, here we have set $\beta_{0} =1$. Thus, our comoving transverse velocity is given by,
\bqe
\beta_{\perp} = \frac{1}{28}. \label{beta}
\eqe   
What is intriguing about this transverse velocity is that it is universal for synchrotron systems with hyperbolic recoil, i.e. it is independent of the beam parameters. The lab frame velocity does indeed depend on the beam energy via $\beta_{\perp lab} = \frac{1}{28 \gamma}$. As such, our differential heat flux is given by, $dQ = \frac{d(\hbar \omega)}{28}$. Let us now develop our photon spectrum based on the second law.

\subsection{The Rindler entropy photon spectrum}

Although it is commonly believed that thermal radiation contains no information, there is, in fact, a persistence of information in the particles radiated \cite{2016PhLB..757..383A}. One intriguing aspect of this is that it sheds light on a potential resolution of the black hole information loss paradox. For the case of black holes, each photon emitted will carry away a portion of the information content which has ``disappeared" behind the event horizon \cite{2018PhLB..776...10A}. This connection between photon number and entropy thereby provides us with a method to link the spectrum of a thermal system to its information content via the second law of thermodynamics.

We begin by exploring the information content carried away by photons which subsequently pass through the Rindler horizon and thereby generate an area change. In consideration of the entropy, $S$, we will then have $\alpha_S$ nats of information per $N$ photons, i.e. $S = \alpha_{S} N$ \cite{2016PhLB..757..383A}, and therefore $dS = \alpha_S dN$, with  $\alpha_{S} = 2.7 \pm 1.7$. From the previous section, we take the differential energy flux that crosses the horizon to be $dQ = \beta_{\perp}d ( \hbar \omega)$. As such, our second law can be written as
\bqe
\frac{dN}{d(\hbar \omega)} = \frac{\beta_{\perp}}{\alpha_{S}} \frac{1}{k_{B}T}.
\eqe
In the above spectrum, the photons will be thermalized at the FDU temperature, $T_{FDU} = \frac{a \hbar }{2 \pi c k_{B}}$. Finally, we note the fact that this is an ultra-relativistic system that is thermalized and thus gives us insight into how a temperature boosts \cite{einstein.., 1966Natur.212..571L, 1967Natur.214..903L, 2017NatSR...717657F}. \textit{We find that the temperature must necessarily boost like a length, i.e.}, $T_{lab} = T_{proper}/\gamma$, in agreement with previous results \cite{2020PhRvD.102h5006B}. As such, in order to match the data set, our temperature must be boosted into the lab frame via, $T_{lab} = T_{FDU}/\gamma$. Thus, \textit{the second law of thermodynamics can be mapped to a particle spectrum}, 
\bqe
\frac{dN}{d(\hbar \omega)} = \frac{1}{28 \alpha_{S}}  \frac{2\pi mc^2 \gamma	}{(\hbar \omega)^2}. \label{spec}
\eqe
What is important to note, is that the entropy content, $\alpha_{S} = 2.7$, is for a 3-dimensional blackbody and its value depends crucially on the dimensionality of the system \cite{1989JPhA...22.1073L}. An important concept to consider is that the dimensionality which governs the thermodynamics of the Rindler horizon may be 1 or perhaps 2-dimensional \cite{Bekenstein:2001tj, 2024PTEP.2024b3D01L}. Moreover, the fact that we are able to examine the entropy content of the photons which pass through the Rindler horizon may give us a platform to experimentally explore processes related to black hole evaporation and the information loss paradox \cite{ 2016PhLB..757..383A, 2018PhLB..776...10A}.

\subsection{Experimental methods}
The CERN-NA63 experimental site studies various aspects of strong field QED \cite{2019PhRvR...1c3014W, 2023PhRvL.130g1601N}. Here we analyse their high energy channeling radiation experiments which successfully measured radiation reaction \cite{2019PhRvR...1c3014W, 2018NatCo...9..795W}. There, an ultra-relativistic $\sim 100$ GeV positrons traverses samples of single crystal silicon along the $\braket{111}$ axis. These ``channeled" positrons are repelled by the atomic lattice and undergo a transverse harmonic oscillation which causes a photon to build up around the positron as it is pumped up the ladder of harmonic oscillator states. This photon becomes so energetic that its energy is comparable to the positron rest mass and upon emission, the positron experiences an enormous recoil acceleration. This acceleration is sufficiently strong enough to thermalize the system via the Unruh effect \cite{lynch2021experimental,2024PhRvD.109j5009L, 2025arXiv250521292L}.

In order to analyze the NA63 data set using the second law, we must first transform their power spectrum, $\frac{dE}{d(\hbar \omega)dt}$ into a photon spectrum, $\frac{dN}{d(\hbar \omega)}$. We shall take the crystal crossing time to be the time interval, $dt = (\Delta x)/c$, where $\Delta x$ is the crystal thickness. Next, we must convert the resultant spectrum, $\frac{dE}{d(\hbar \omega)}$ into a probability density, i.e. histogram, by dividing by the bin width, $\frac{dE}{d(\hbar \omega)} \frac{1}{\Delta E_{b}}$, where $\Delta E_{b} =1.007$ GeV is the bin width for the 178.2 GeV samples and $\Delta E_{b} =1$ GeV for the 50 GeV samples, and then normalizing the data. \textit{It is this differential bin spectrum which we will analyze.} 

To compare our second law spectrum to the data, we present 4 theoretical spectra for each data set. Since all parameters are completely fixed, we show the plots with $\alpha_{S} = 2.7$ and $\alpha_{S} = 2.7 \pm \sigma_S$, with $\sigma_S = 1.7$. We also include a plot with the best fit entropy per photon, $\alpha^{*}_{S}$. For all data sets we find $2.7 - 1.7< \alpha^{*}_{S} < 2.7 + 1.7$. For the best fit $\alpha^{*}_{S}$ we also include a reduced chi-squared statistic, $\chi^{2}/\nu = \frac{\lb y_{theory} - y_{exp} \rb^2}{\Delta y_{exp} \nu}$. Here, $\Delta y_{exp}$ is the error and $\nu = \omega_{max} - \omega^{*} - 1$ is the number of degrees of freedom. Here, $\omega^{*} = \omega_{c}+E/10$. This way, our chi-squared and best fit $\alpha^{*}_{S}$ are determined above the threshold of $\omega_c$ to ensure we are well within the regime where the Unruh effect dominates. In all cases, we find the best fit spectra reside within the bands determined by the entropy per photon uncertainty. Moreover, we find that the high frequency tail has a $\sim \frac{1}{( \hbar \omega)^2}$ dependence which corroborates the fact that the data sets provide \textit{a direct measurement of the FDU temperature} across the spectrum of recoil. A comparison of the second law spectrum, Eqn. (\ref{spec}), to each data set is presented below.	

\subsection{Second law spectra comparison}

\begin{figure}[H]
\centering  
\includegraphics[scale=.27]{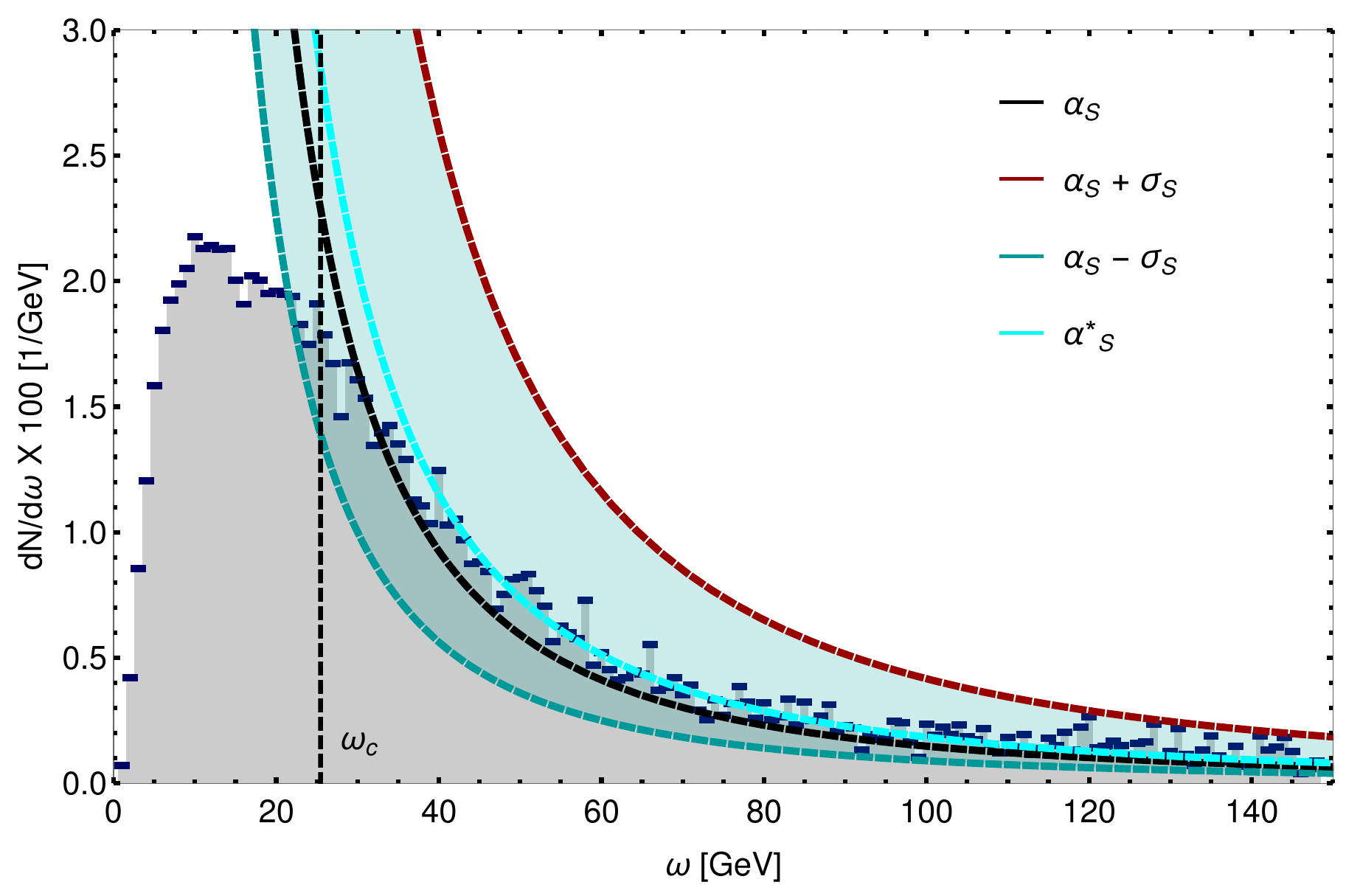}
\caption{The second law spectrum for the 178.2 GeV, 3.8 mm crystal with a best fit $\alpha^{*}_S = 2.17$ and $\chi^2/\nu = 1.11$.} 	
\label{plot38bit}
\end{figure}

\begin{figure}[H]
\centering  
\includegraphics[scale=.27]{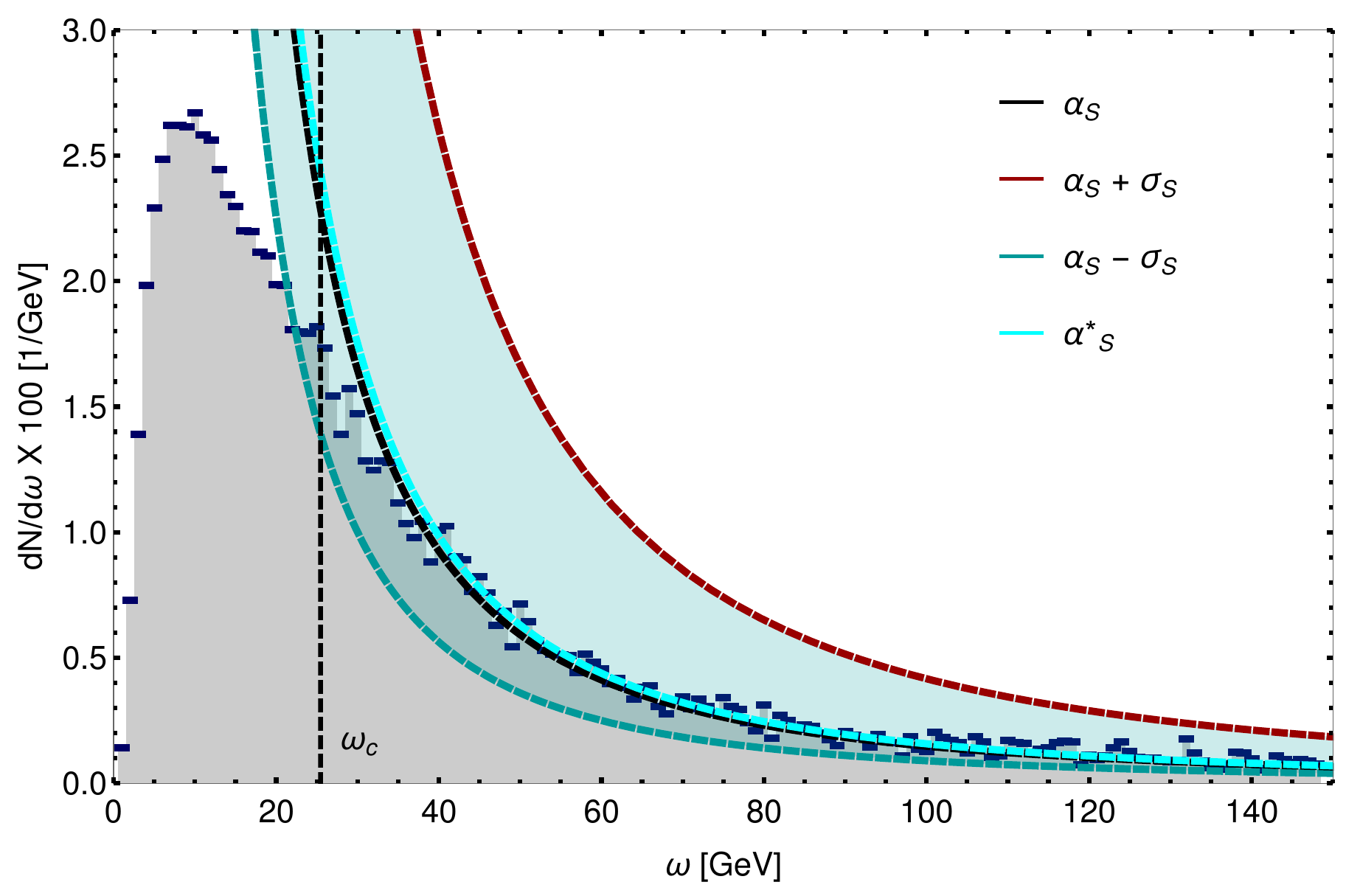}
\caption{The second law spectrum for the 178.2 GeV, 10 mm crystal with a best fit $\alpha^{*}_S = 2.54$ and $\chi^2/\nu = 1.36$.} 	
\label{10bit}
\end{figure}

\begin{figure}[H]
\centering  
\includegraphics[scale=.27]{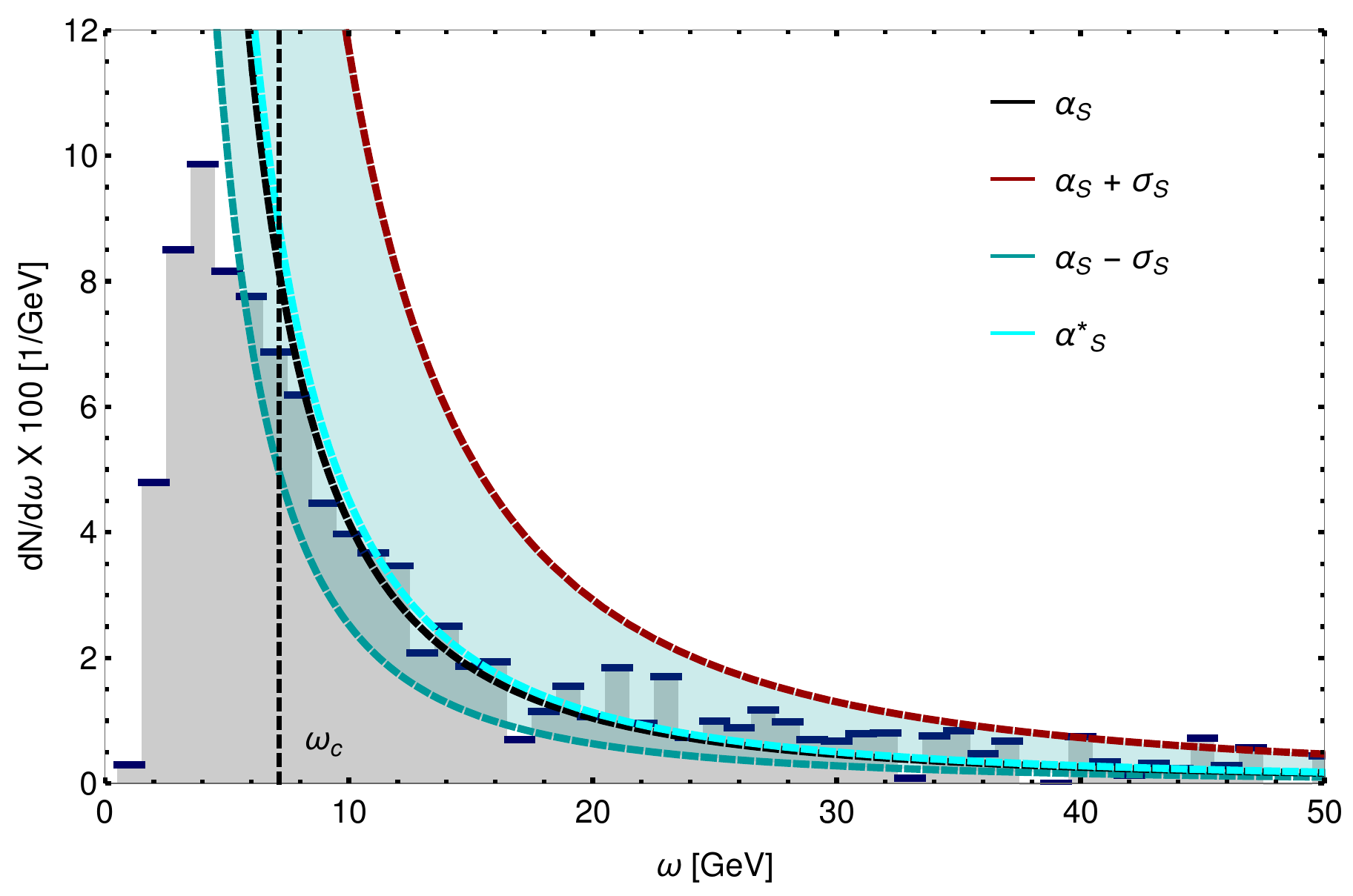}
\caption{The second law spectrum for the 50 GeV, 1 mm crystal with a best fit $\alpha^{*}_S = 2.48$ and $\chi^2/\nu = 2.91$.} 	
\label{1bit}
\end{figure}

\begin{figure}[H]
\centering  
\includegraphics[scale=.27]{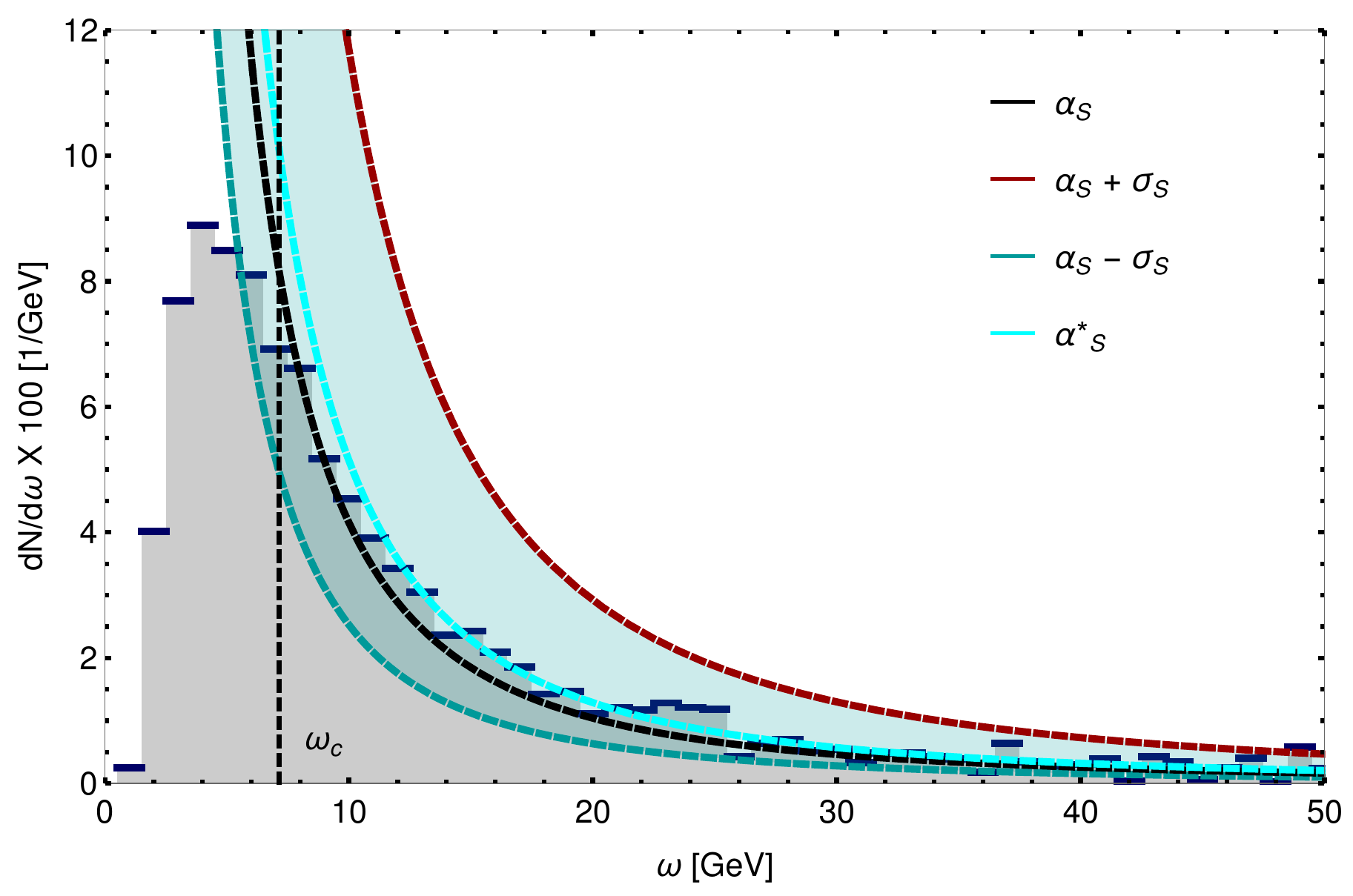}
\caption{The second law spectrum for the 50 GeV, 2 mm crystal with a best fit $\alpha^{*}_S = 2.18$ and $\chi^2/\nu = 2.15$.} 	
\label{2bit}
\end{figure}

\begin{figure}[H]
\centering  
\includegraphics[scale=.27]{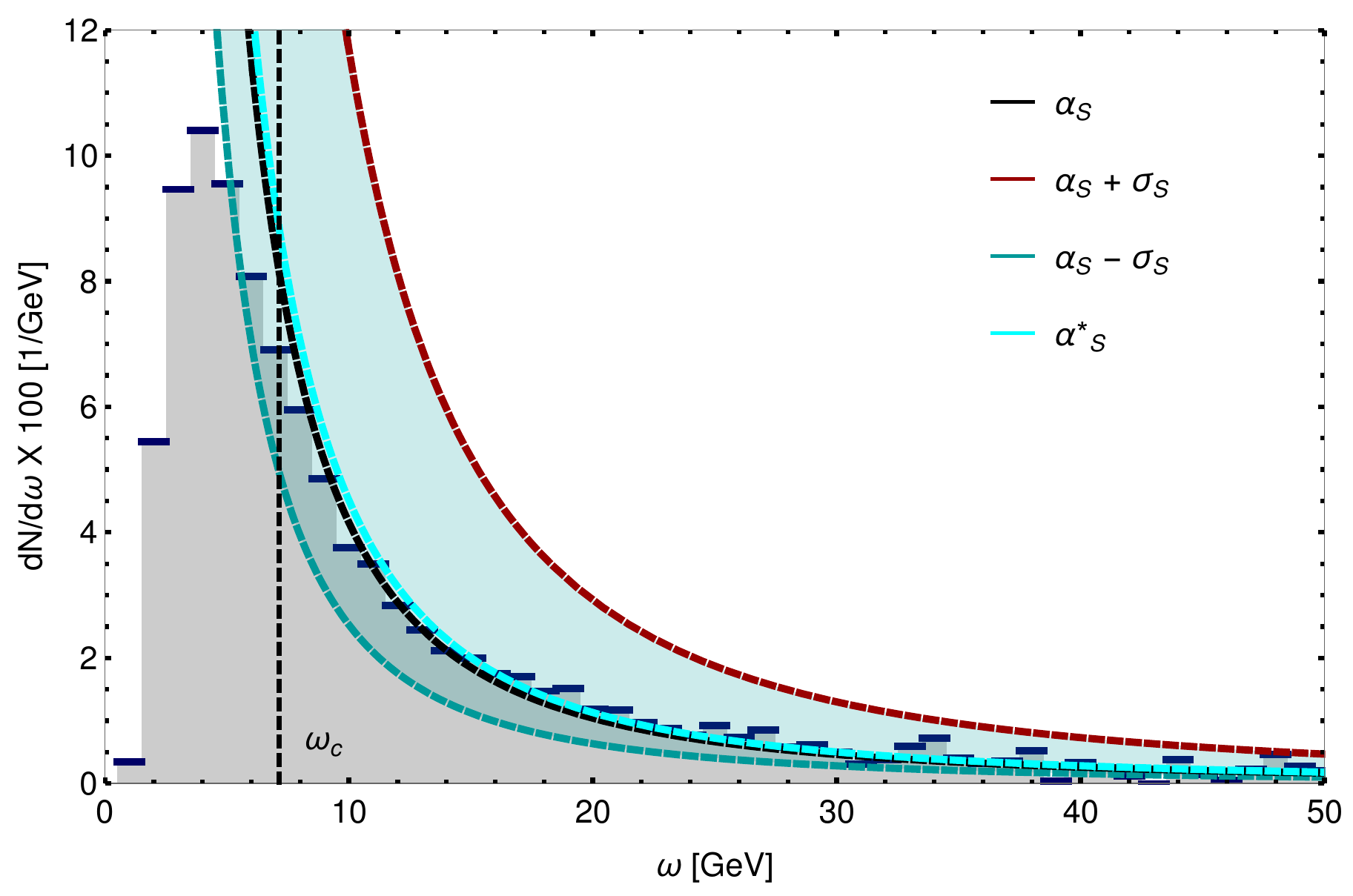}
\caption{The second law spectrum for the 50 GeV, 4 mm crystal with a best fit $\alpha^{*}_S = 2.49$ and $\chi^2/\nu = 1.53$.} 	
\label{4bit}
\end{figure}

\begin{figure}[H]
\centering  
\includegraphics[scale=.27]{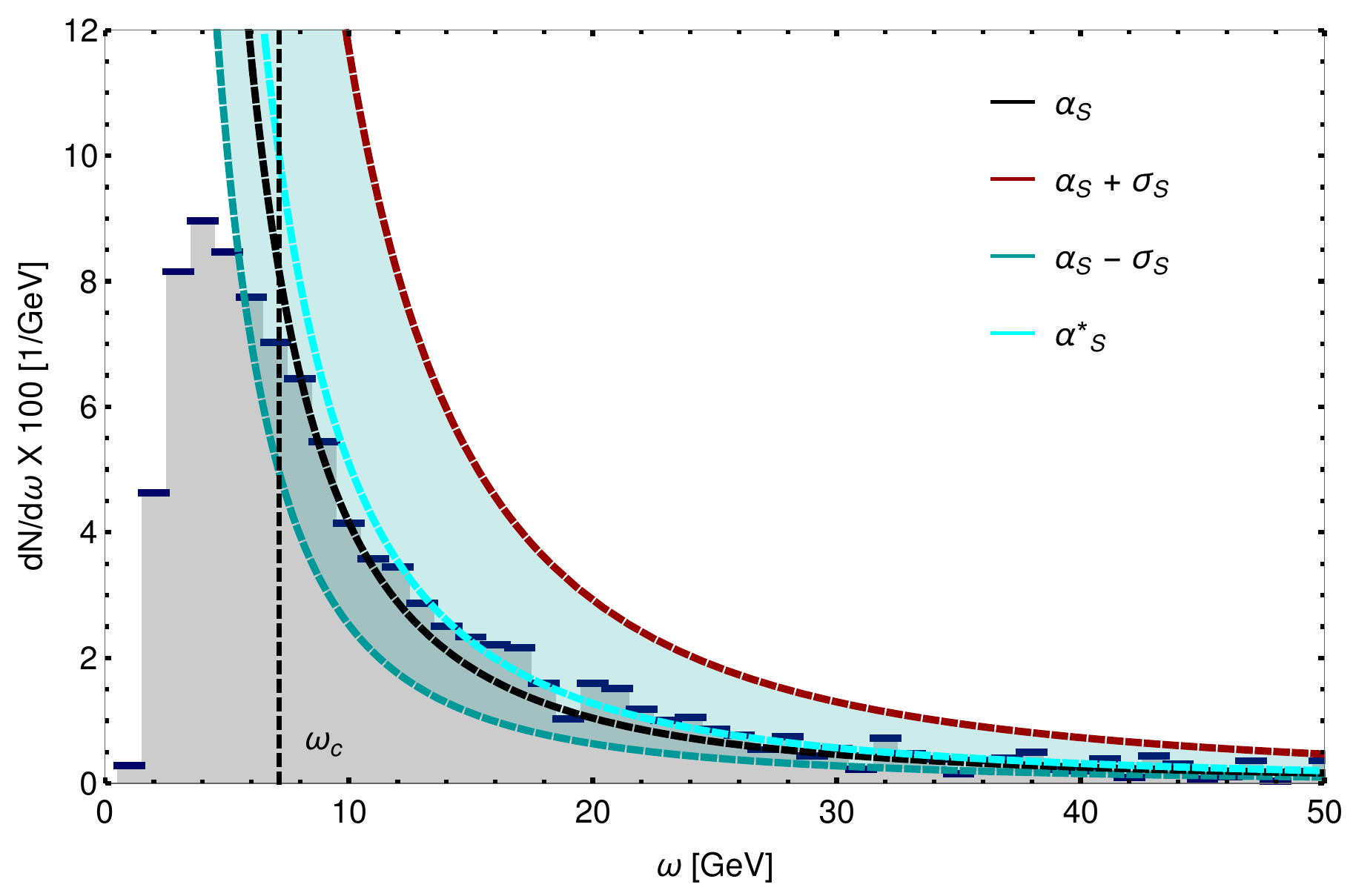}
\caption{The second law spectrum for the 50 GeV, 6-1 mm crystal with a best fit $\alpha^{*}_S = 2.20$ and $\chi^2/\nu = 1.39$.} 	
\label{61bit}
\end{figure}

\begin{figure}[H]
\centering  
\includegraphics[scale=.27]{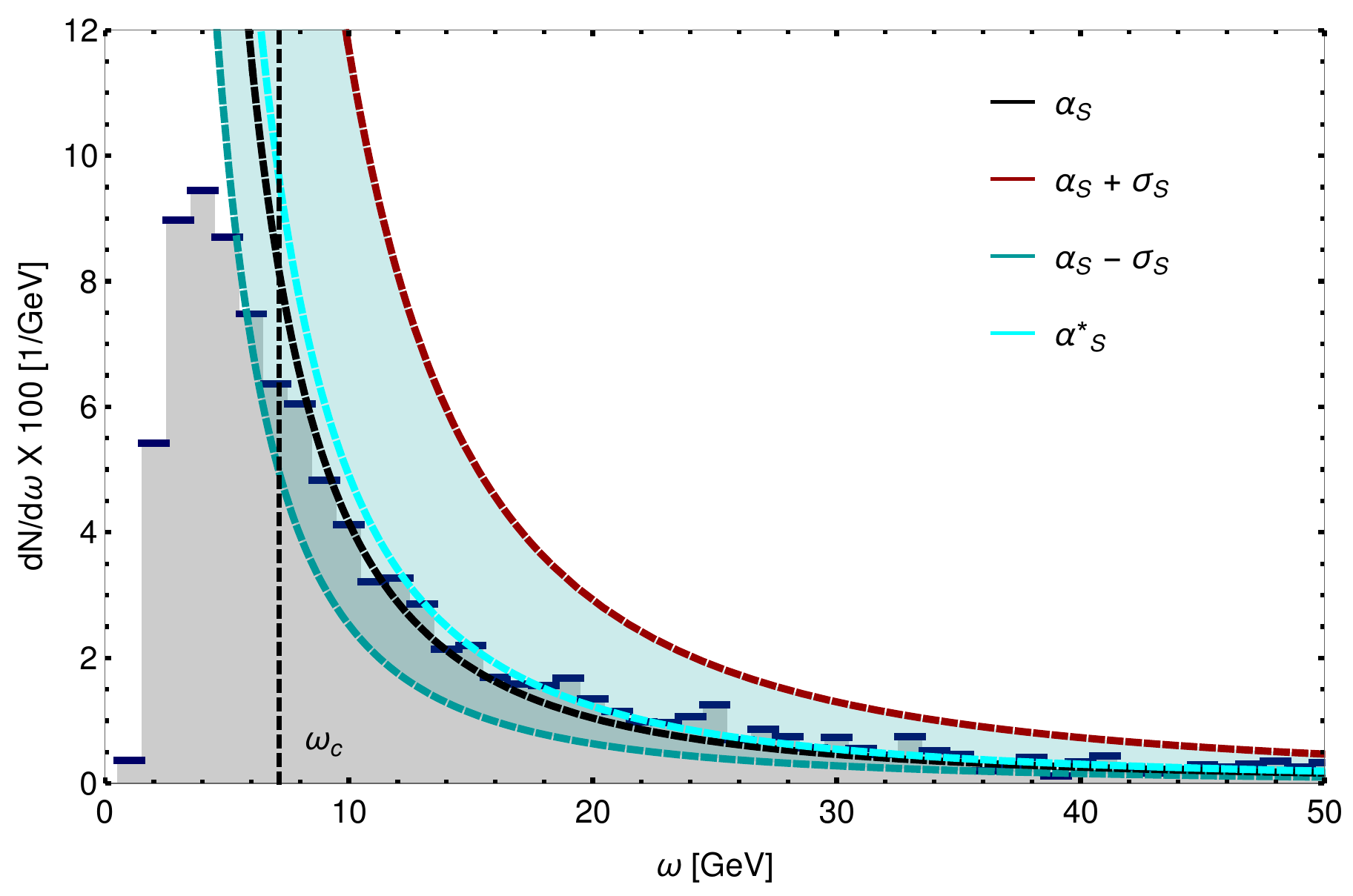}
\caption{The second law spectrum for the 50 GeV, 6-2 mm crystal with a best fit $\alpha^{*}_S = 2.29$ and $\chi^2/\nu = 1.33$.} 	
\label{62bit}
\end{figure}

\subsection{Measurement of the positron mass}
Given a successful description of the photon spectrum based on the second law of thermodynamics, it is interesting to note that within the temperature, one is able to isolate any of the physical constants present; namely $c$, $k_{B}$, $\hbar$, or the positron mass $m_{p^{+}}$. This provides an interesting technique for measuring physical constants. Although many of these constants are used as experimental inputs prior to the measurement, it seems reasonable that one, say the positron mass, $m_{p^{+}}$,  may be left out of any preliminary experimental inputs and then be measured via the overall coefficient on the frequency dependent temperature, $T_{FDU}$. Thus, we can solve for the positron mass to yield,
\bqe
m_{p^{+}} = 28\alpha_{S}^{*} \frac{(\hbar \omega)^2	}{2 \pi c^2\gamma}\frac{dN_{EXP}}{d(\hbar \omega)}.
\eqe
As such, we find, via the recoil FDU temperature, a thermometer capable of measuring mass and/or other physical constants. The measurements of the positron mass are presented below. From the data sets, we average the measured positron mass over the region, $\omega^* \rightarrow E$ for each sample. The measured mass of each sample is presented in the plots below. Then, we average over each of the measured masses to obtain a value of $m_{p^+}= 10.0 \pm 1.63 \times 10^{-30}$ kg or $.562 \pm .091$ MeV/$c^2$. The error is computed via $\delta m_{tot} =\frac{1}{N}\sum^{N}_{i}\sqrt{(\delta m)^2}$, where $\delta m$ is the standard deviation of each individual mass measurement. Note, the statistical error here is the dominant source of error as the experimental error bars are smaller than the data points. The accepted value of the positron mass is $m_{p} = 9.11 \times 10^{-31}$ kg or $.511$ MeV/$c^{2}$ \cite{Workman:2022ynf}. \\

\begin{figure}[H]
\centering  
\includegraphics[scale=.27]{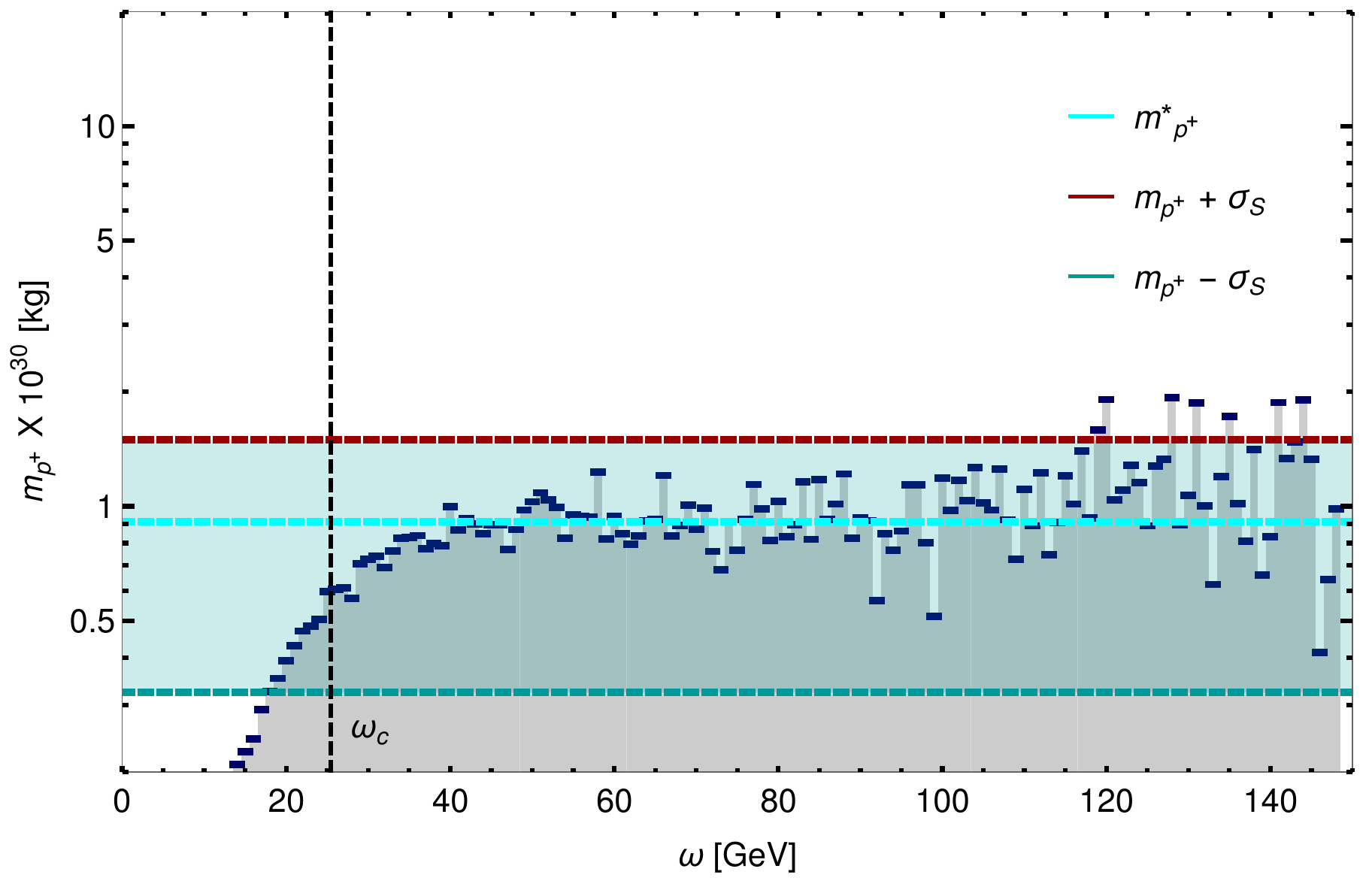}
\caption{The positron mass measurement for the 178.2 GeV, 3.8 mm crystal with $m_{p^{+}} = 10.2 \pm 2.85 \times 10^{-31}$ kg.} 	
\label{38mass}
\end{figure}

\begin{figure}[H]
\centering  
\includegraphics[scale=.27]{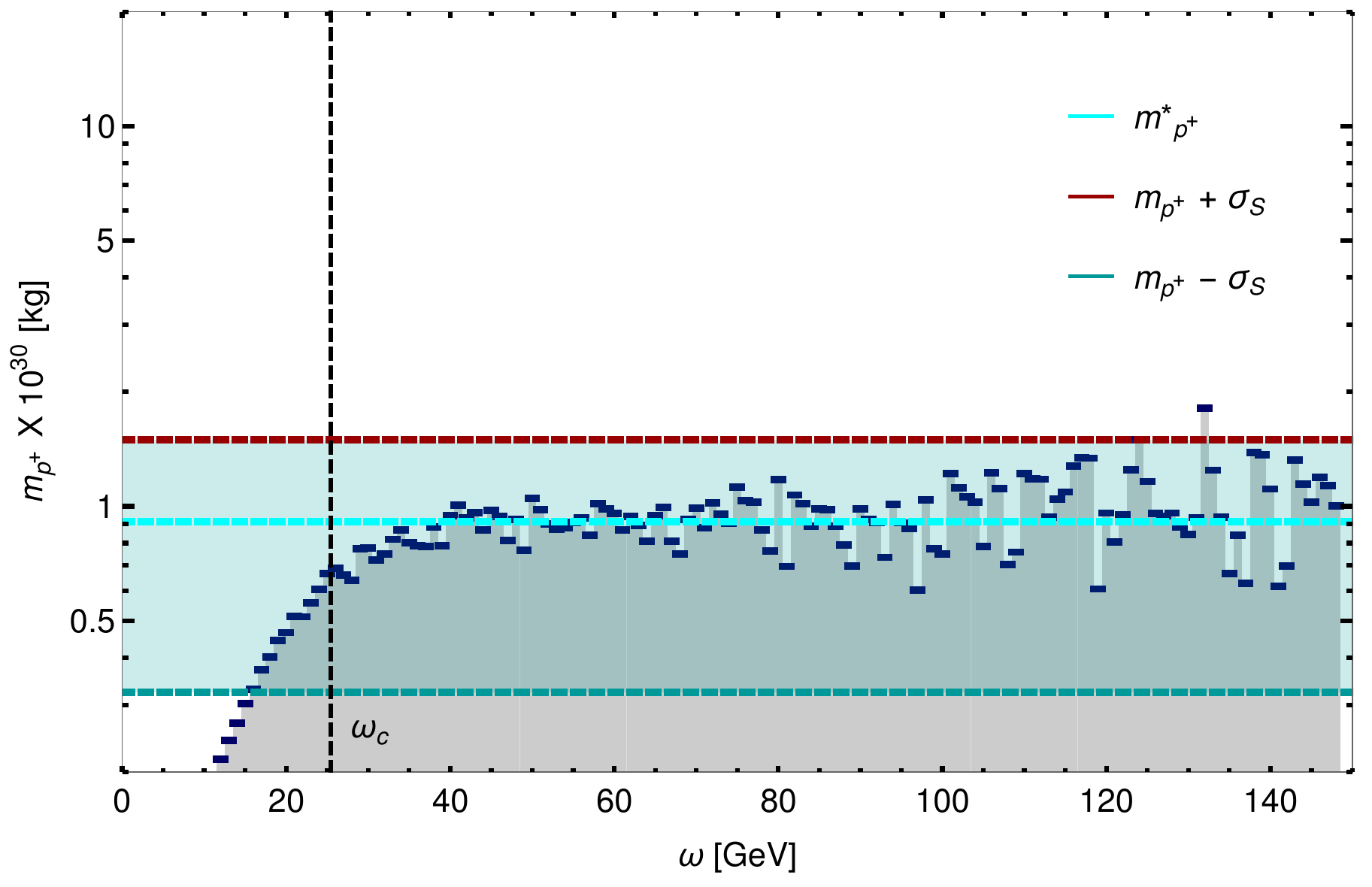}
\caption{The positron mass measurement for the 178.2 GeV, 10 mm crystal with $m_{p^{+}} = 9.73 \pm 1.96 \times 10^{-31}$ kg.} 	
\label{10mass}
\end{figure}

\begin{figure}[H]
\centering  
\includegraphics[scale=.27]{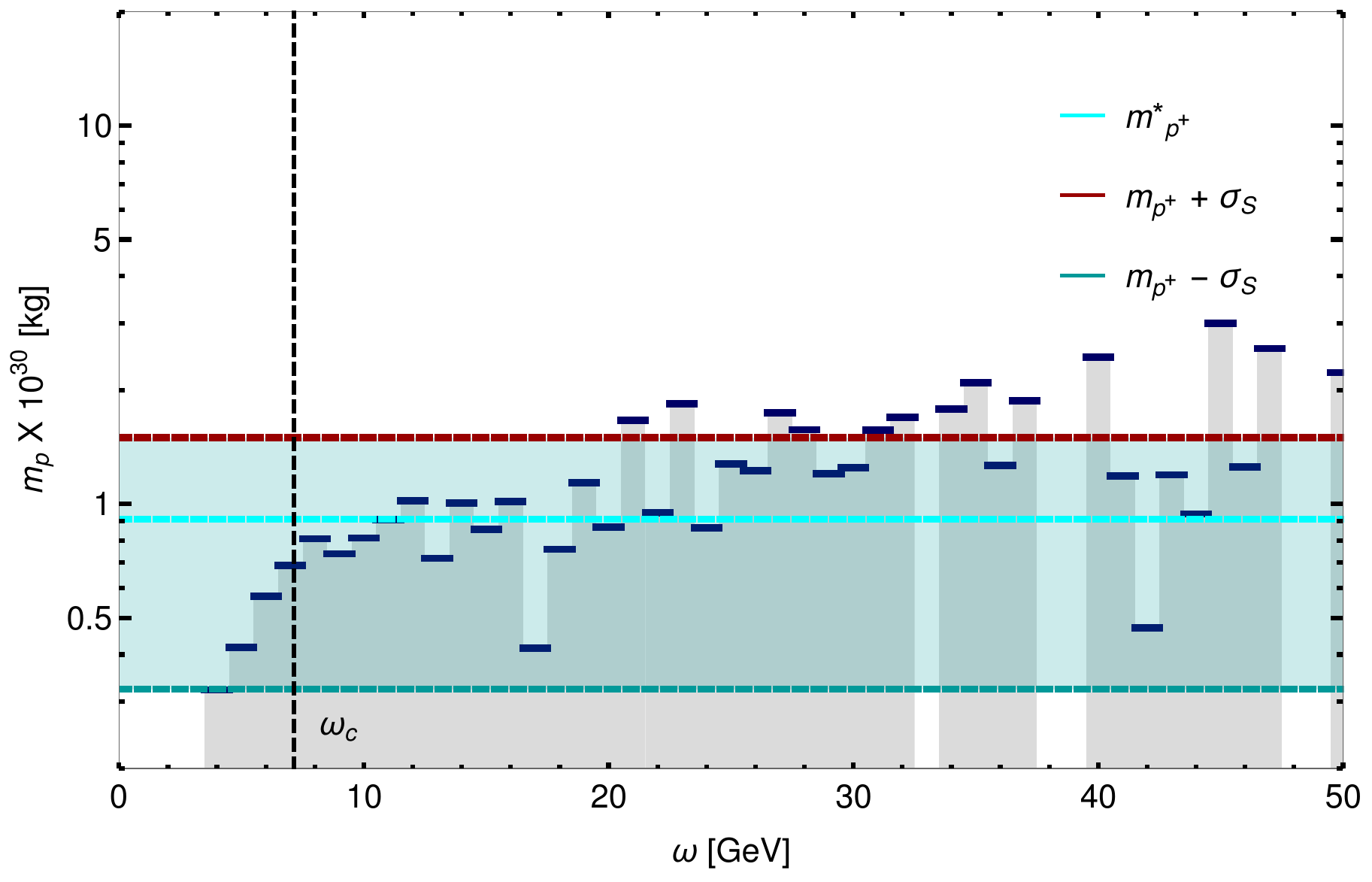}
\caption{The positron mass measurement for the 50 GeV, 1 mm crystal with $m_{p^{+}} = 11.7 \pm 8.11 \times 10^{-31}$ kg.} 	
\label{1mass}
\end{figure}

\begin{figure}[H]
\centering  
\includegraphics[scale=.27]{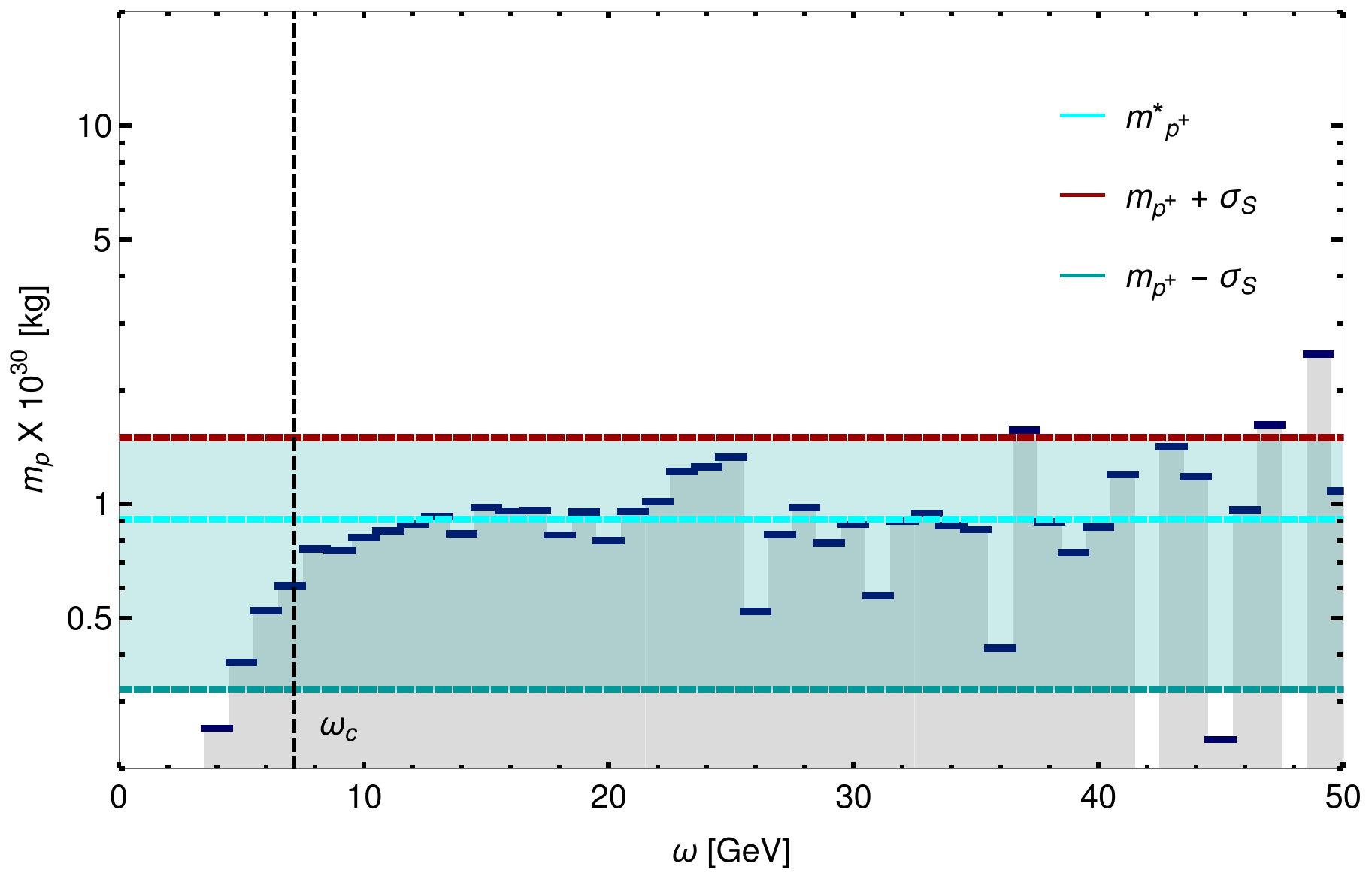}
\caption{The positron mass measurement for the 50 GeV, 2 mm crystal with $m_{p^{+}} = 9.49 \pm 4.14 \times 10^{-31}$ kg.} 	
\label{2mass}
\end{figure}

\begin{figure}[H]
\centering  
\includegraphics[scale=.27]{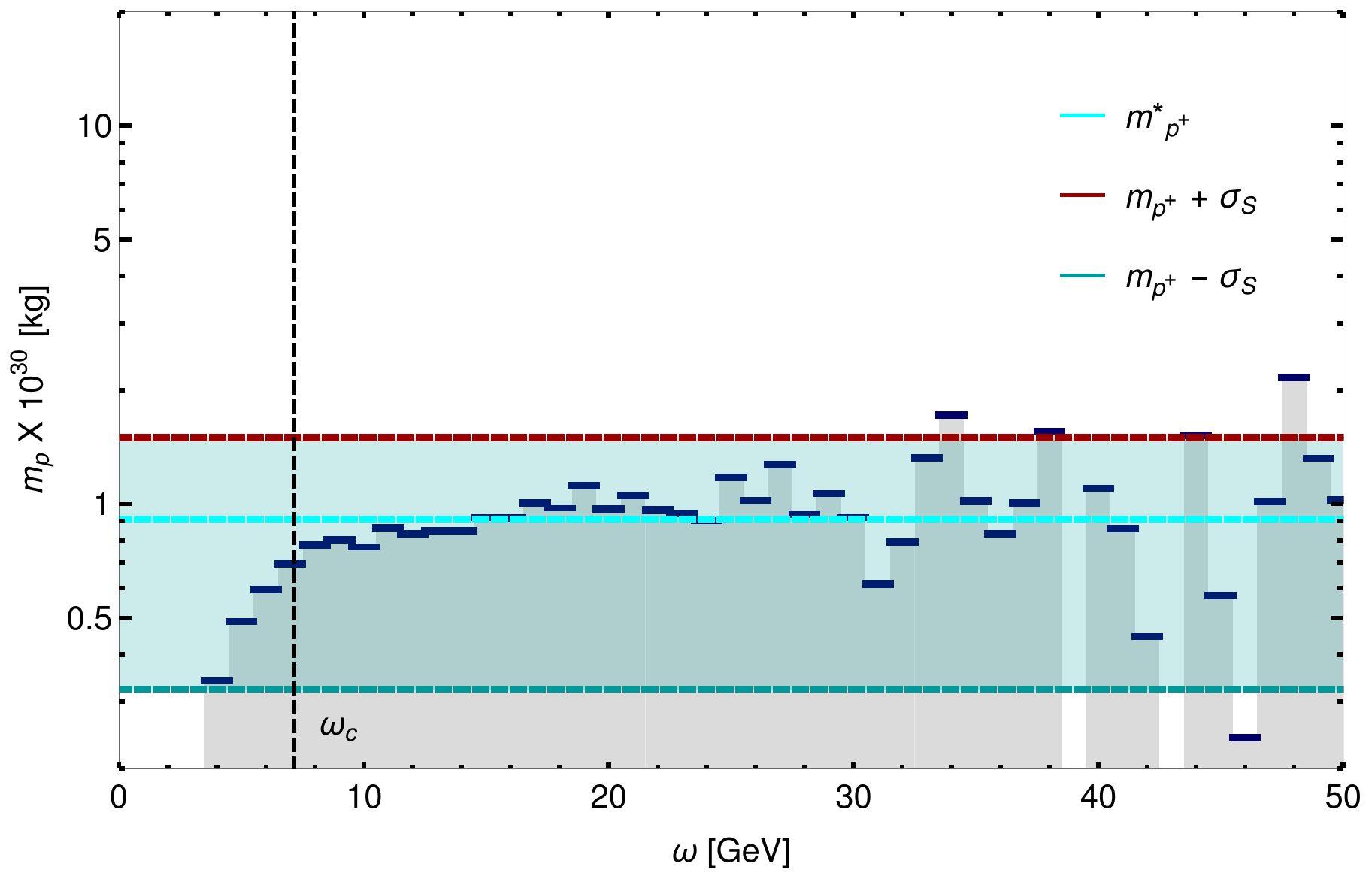}
\caption{The positron mass measurement for the 50 GeV, 4 mm crystal with $m_{p^{+}} = 9.75 \pm 3.99 \times 10^{-31}$ kg.} 	
\label{4mass}
\end{figure}

\begin{figure}[H]
\centering  
\includegraphics[scale=.27]{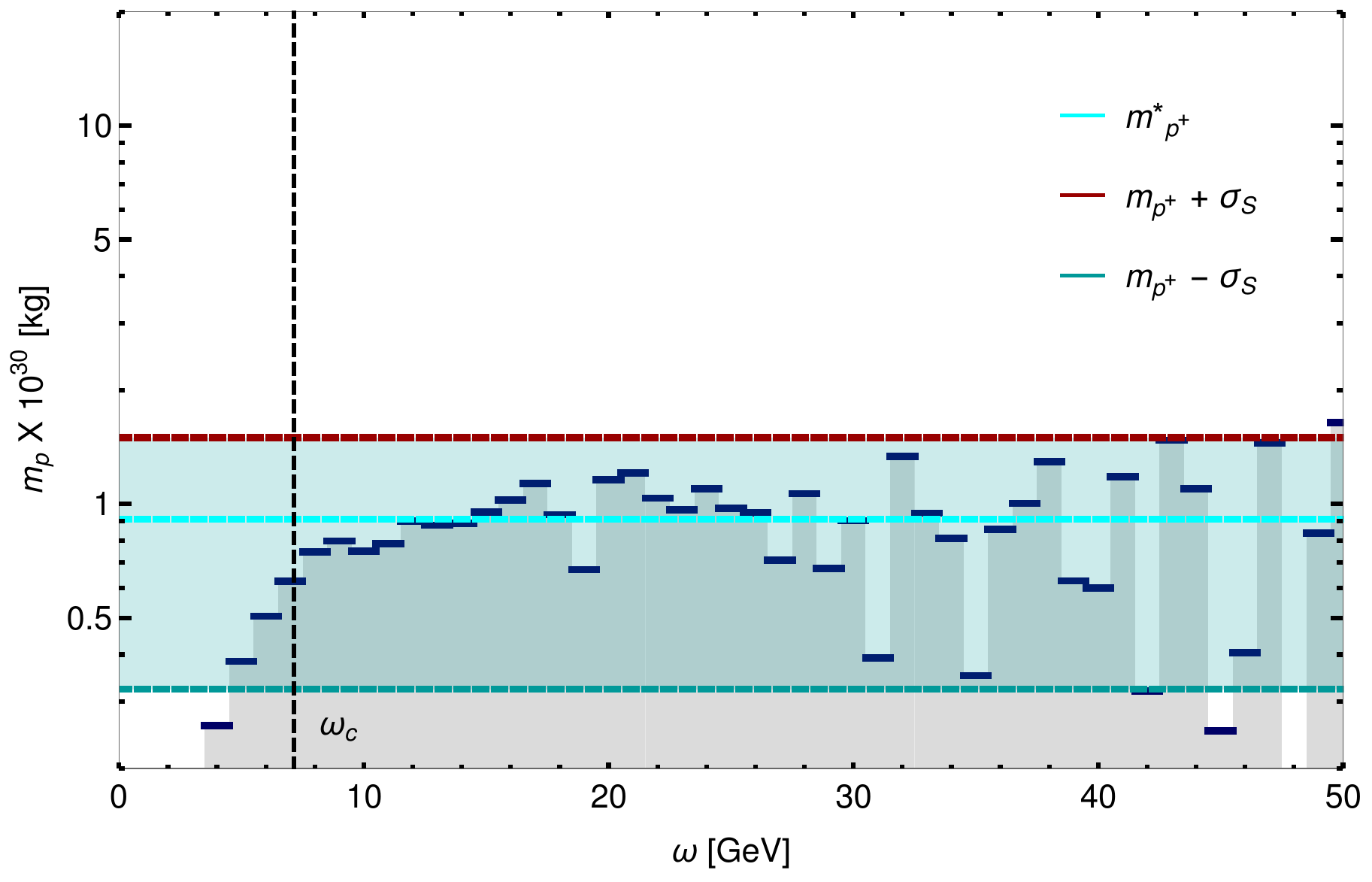}
\caption{The positron mass measurement for the 50 GeV, 6-1 mm crystal with $m_{p^{+}} = 9.01 \pm 3.41 \times 10^{-31}$ kg.} 	
\label{61mass}
\end{figure}

\begin{figure}[H]
\centering  
\includegraphics[scale=.27]{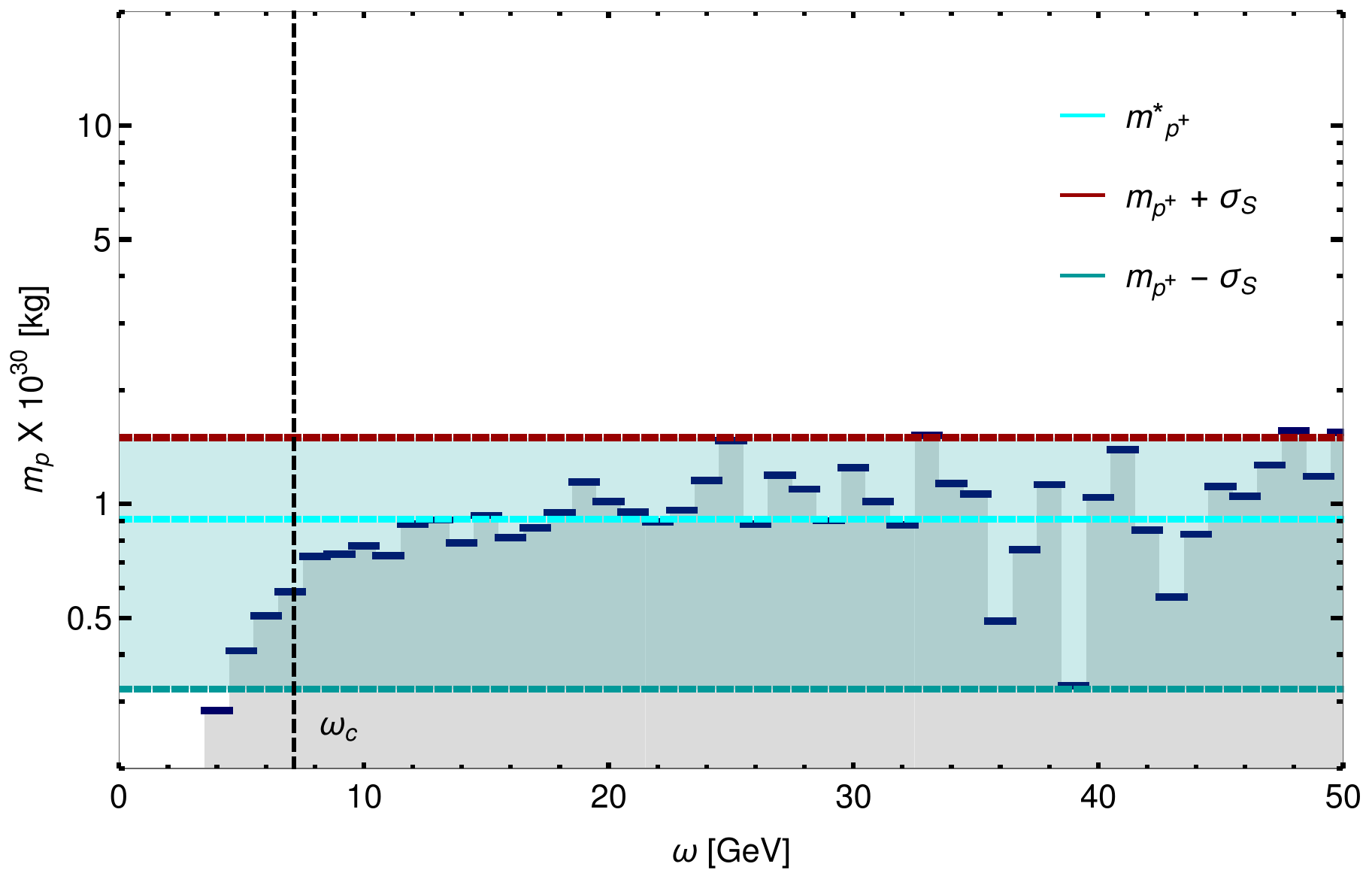}
\caption{The positron mass measurement for the 50 GeV, 4 mm crystal with $m_{p^{+}} = 10.2 \pm 2.67 \times 10^{-31}$ kg.} 	
\label{62mass}
\end{figure}

\section{conclusions}
In this manuscript, we have analyzed the Unruh-thermalized photon spectrum of channeled positrons undergoing the extreme accelerations of radiation reaction measured by CERN-NA63. We employed an analysis based entirely on the 2nd law of Rindler horizon thermodynamics, at the recoil FDU temperature, and found an excellent agreement with a suite of 7 data sets. In particular, we found that the Unruh effect saturates the spectra beyond the critical frequency determined by an asymptotic timescale analysis. Moreover, the ultra-relativistic nature of the system reveals that temperatures boost like a length, i.e. $T_{lab} = T_{proper}/\gamma$. This thermodynamic technique also provides a novel way to experimentally measure the positron mass. 

\section*{Acknowledgments}
The author thanks the Seoul National University for their hospitality during the development of this manuscript. This work has been supported by the National Research Foundation of Korea under Grants No.~2017R1A2A2A05001422 and No.~2020R1A2C2008103.

\goodbreak

\bibliography{temp}

\begin{thebibliography}{30}%
\makeatletter
\providecommand \@ifxundefined [1]{%
 \@ifx{#1\undefined}
}%
\providecommand \@ifnum [1]{%
 \ifnum #1\expandafter \@firstoftwo
 \else \expandafter \@secondoftwo
 \fi
}%
\providecommand \@ifx [1]{%
 \ifx #1\expandafter \@firstoftwo
 \else \expandafter \@secondoftwo
 \fi
}%
\providecommand \natexlab [1]{#1}%
\providecommand \enquote  [1]{``#1''}%
\providecommand \bibnamefont  [1]{#1}%
\providecommand \bibfnamefont [1]{#1}%
\providecommand \citenamefont [1]{#1}%
\providecommand \href@noop [0]{\@secondoftwo}%
\providecommand \href [0]{\begingroup \@sanitize@url \@href}%
\providecommand \@href[1]{\@@startlink{#1}\@@href}%
\providecommand \@@href[1]{\endgroup#1\@@endlink}%
\providecommand \@sanitize@url [0]{\catcode `\\12\catcode `\$12\catcode
  `\&12\catcode `\#12\catcode `\^12\catcode `\_12\catcode `\%12\relax}%
\providecommand \@@startlink[1]{}%
\providecommand \@@endlink[0]{}%
\providecommand \url  [0]{\begingroup\@sanitize@url \@url }%
\providecommand \@url [1]{\endgroup\@href {#1}{\urlprefix }}%
\providecommand \urlprefix  [0]{URL }%
\providecommand \Eprint [0]{\href }%
\providecommand \doibase [0]{http://dx.doi.org/}%
\providecommand \selectlanguage [0]{\@gobble}%
\providecommand \bibinfo  [0]{\@secondoftwo}%
\providecommand \bibfield  [0]{\@secondoftwo}%
\providecommand \translation [1]{[#1]}%
\providecommand \BibitemOpen [0]{}%
\providecommand \bibitemStop [0]{}%
\providecommand \bibitemNoStop [0]{.\EOS\space}%
\providecommand \EOS [0]{\spacefactor3000\relax}%
\providecommand \BibitemShut  [1]{\csname bibitem#1\endcsname}%
\let\auto@bib@innerbib\@empty
\bibitem [{\citenamefont {Parker}(1968)}]{Parker:1968mv}%
  \BibitemOpen
  \bibfield  {author} {\bibinfo {author} {\bibfnamefont {L.}~\bibnamefont
  {Parker}},\ }\bibfield  {title} {\enquote {\bibinfo {title} {{Particle
  creation in expanding universes}},}\ }\href {\doibase
  10.1103/PhysRevLett.21.562} {\bibfield  {journal} {\bibinfo  {journal} {Phys.
  Rev. Lett.}\ }\textbf {\bibinfo {volume} {21}},\ \bibinfo {pages} {562--564}
  (\bibinfo {year} {1968})}\BibitemShut {NoStop}%
\bibitem [{\citenamefont {Hawking}(1974)}]{hawking1974black}%
  \BibitemOpen
  \bibfield  {author} {\bibinfo {author} {\bibfnamefont {Stephen~W}\
  \bibnamefont {Hawking}},\ }\bibfield  {title} {\enquote {\bibinfo {title}
  {Black hole explosions?}}\ }\href {\doibase https://doi.org/10.1038/248030a0}
  {\bibfield  {journal} {\bibinfo  {journal} {Nature}\ }\textbf {\bibinfo
  {volume} {248}},\ \bibinfo {pages} {30--31} (\bibinfo {year}
  {1974})}\BibitemShut {NoStop}%
\bibitem [{\citenamefont {Fulling}\ and\ \citenamefont
  {Davies}(1976)}]{Davies:1976hi}%
  \BibitemOpen
  \bibfield  {author} {\bibinfo {author} {\bibfnamefont {S.~A.}\ \bibnamefont
  {Fulling}}\ and\ \bibinfo {author} {\bibfnamefont {P.~C.~W.}\ \bibnamefont
  {Davies}},\ }\bibfield  {title} {\enquote {\bibinfo {title} {Radiation from a
  moving mirror in two dimensional space-time: conformal anomaly},}\ }\href
  {https://royalsocietypublishing.org/doi/abs/10.1098/rspa.1976.0045}
  {\bibfield  {journal} {\bibinfo  {journal} {Proc. R. Soc. Lond. A}\ }\textbf
  {\bibinfo {volume} {348}},\ \bibinfo {pages} {393--414} (\bibinfo {year}
  {1976})}\BibitemShut {NoStop}%
\bibitem [{\citenamefont {Davies}\ and\ \citenamefont
  {Fulling}(1977)}]{Davies:1977yv}%
  \BibitemOpen
  \bibfield  {author} {\bibinfo {author} {\bibfnamefont {P.C.W.}\ \bibnamefont
  {Davies}}\ and\ \bibinfo {author} {\bibfnamefont {S.A.}\ \bibnamefont
  {Fulling}},\ }\bibfield  {title} {\enquote {\bibinfo {title} {{Radiation from
  Moving Mirrors and from Black Holes}},}\ }\href {\doibase
  10.1098/rspa.1977.0130} {\bibfield  {journal} {\bibinfo  {journal} {Proc. R.
  Soc. Lond. A}\ }\textbf {\bibinfo {volume} {A356}},\ \bibinfo {pages}
  {237--257} (\bibinfo {year} {1977})}\BibitemShut {NoStop}%
\bibitem [{\citenamefont {Unruh}(1976)}]{Unruh:1976db}%
  \BibitemOpen
  \bibfield  {author} {\bibinfo {author} {\bibfnamefont {W.~G.}\ \bibnamefont
  {Unruh}},\ }\bibfield  {title} {\enquote {\bibinfo {title} {{Notes on black
  hole evaporation}},}\ }\href {\doibase 10.1103/PhysRevD.14.870} {\bibfield
  {journal} {\bibinfo  {journal} {Phys. Rev. D}\ }\textbf {\bibinfo {volume}
  {14}},\ \bibinfo {pages} {870} (\bibinfo {year} {1976})}\BibitemShut
  {NoStop}%
\bibitem [{\citenamefont {Lynch}\ \emph {et~al.}(2021)\citenamefont {Lynch},
  \citenamefont {Cohen}, \citenamefont {Hadad},\ and\ \citenamefont
  {Kaminer}}]{lynch2021experimental}%
  \BibitemOpen
  \bibfield  {author} {\bibinfo {author} {\bibfnamefont {Morgan~H}\
  \bibnamefont {Lynch}}, \bibinfo {author} {\bibfnamefont {Eliahu}\
  \bibnamefont {Cohen}}, \bibinfo {author} {\bibfnamefont {Yaron}\ \bibnamefont
  {Hadad}}, \ and\ \bibinfo {author} {\bibfnamefont {Ido}\ \bibnamefont
  {Kaminer}},\ }\bibfield  {title} {\enquote {\bibinfo {title} {Experimental
  observation of acceleration-induced thermality},}\ }\href {\doibase
  https://doi.org/10.1103/PhysRevD.104.025015} {\bibfield  {journal} {\bibinfo
  {journal} {Physical Review D}\ }\textbf {\bibinfo {volume} {104}},\ \bibinfo
  {pages} {025015} (\bibinfo {year} {2021})},\ \Eprint
  {http://arxiv.org/abs/1903.00043} {arXiv:1903.00043 [gr-qc]} \BibitemShut
  {NoStop}%
\bibitem [{\citenamefont {{Lynch}}\ \emph {et~al.}(2024)\citenamefont
  {{Lynch}}, \citenamefont {{Ievlev}},\ and\ \citenamefont
  {{Good}}}]{2024PTEP.2024b3D01L}%
  \BibitemOpen
  \bibfield  {author} {\bibinfo {author} {\bibfnamefont {Morgan~H.}\
  \bibnamefont {{Lynch}}}, \bibinfo {author} {\bibfnamefont {Evgenii}\
  \bibnamefont {{Ievlev}}}, \ and\ \bibinfo {author} {\bibfnamefont {Michael
  R.~R.}\ \bibnamefont {{Good}}},\ }\bibfield  {title} {\enquote {\bibinfo
  {title} {{Accelerated electron thermometer: observation of 1D Planck
  radiation}},}\ }\href {\doibase 10.1093/ptep/ptad157} {\bibfield  {journal}
  {\bibinfo  {journal} {Progress of Theoretical and Experimental Physics}\
  }\textbf {\bibinfo {volume} {2024}},\ \bibinfo {eid} {023D01} (\bibinfo
  {year} {2024})},\ \Eprint {http://arxiv.org/abs/2211.14774} {arXiv:2211.14774
  [nucl-ex]} \BibitemShut {NoStop}%
\bibitem [{\citenamefont {Bekenstein}(1973)}]{Bekenstein:1973ur}%
  \BibitemOpen
  \bibfield  {author} {\bibinfo {author} {\bibfnamefont {Jacob~D.}\
  \bibnamefont {Bekenstein}},\ }\bibfield  {title} {\enquote {\bibinfo {title}
  {{Black holes and entropy}},}\ }\href {\doibase 10.1103/PhysRevD.7.2333}
  {\bibfield  {journal} {\bibinfo  {journal} {Phys. Rev. D}\ }\textbf {\bibinfo
  {volume} {7}},\ \bibinfo {pages} {2333--2346} (\bibinfo {year}
  {1973})}\BibitemShut {NoStop}%
\bibitem [{\citenamefont {Bianchi}\ and\ \citenamefont
  {Satz}(2013)}]{bianchi2013mechanical}%
  \BibitemOpen
  \bibfield  {author} {\bibinfo {author} {\bibfnamefont {Eugenio}\ \bibnamefont
  {Bianchi}}\ and\ \bibinfo {author} {\bibfnamefont {Alejandro}\ \bibnamefont
  {Satz}},\ }\bibfield  {title} {\enquote {\bibinfo {title} {Mechanical laws of
  the rindler horizon},}\ }\href {\doibase
  https://doi.org/10.1103/PhysRevD.87.124031} {\bibfield  {journal} {\bibinfo
  {journal} {Physical Review D}\ }\textbf {\bibinfo {volume} {87}},\ \bibinfo
  {pages} {124031} (\bibinfo {year} {2013})},\ \Eprint
  {http://arxiv.org/abs/1305.4986} {arXiv:1305.4986 [gr-qc]} \BibitemShut
  {NoStop}%
\bibitem [{\citenamefont {{Jacobson}}(1995)}]{1995PhRvL..75.1260J}%
  \BibitemOpen
  \bibfield  {author} {\bibinfo {author} {\bibfnamefont {Ted}\ \bibnamefont
  {{Jacobson}}},\ }\bibfield  {title} {\enquote {\bibinfo {title}
  {{Thermodynamics of Spacetime: The Einstein Equation of State}},}\ }\href
  {\doibase 10.1103/PhysRevLett.75.1260} {\bibfield  {journal} {\bibinfo
  {journal} {\prl}\ }\textbf {\bibinfo {volume} {75}},\ \bibinfo {pages}
  {1260--1263} (\bibinfo {year} {1995})},\ \Eprint
  {http://arxiv.org/abs/gr-qc/9504004} {arXiv:gr-qc/9504004 [gr-qc]}
  \BibitemShut {NoStop}%
\bibitem [{\citenamefont {{Wistisen}}\ \emph {et~al.}(2018)\citenamefont
  {{Wistisen}}, \citenamefont {{Di Piazza}}, \citenamefont {{Knudsen}},\ and\
  \citenamefont {{Uggerh{\o}j}}}]{2018NatCo...9..795W}%
  \BibitemOpen
  \bibfield  {author} {\bibinfo {author} {\bibfnamefont {Tobias~N.}\
  \bibnamefont {{Wistisen}}}, \bibinfo {author} {\bibfnamefont {Antonino}\
  \bibnamefont {{Di Piazza}}}, \bibinfo {author} {\bibfnamefont {Helge~V.}\
  \bibnamefont {{Knudsen}}}, \ and\ \bibinfo {author} {\bibfnamefont
  {Ulrik~I.}\ \bibnamefont {{Uggerh{\o}j}}},\ }\bibfield  {title} {\enquote
  {\bibinfo {title} {{Experimental evidence of quantum radiation reaction in
  aligned crystals}},}\ }\href {\doibase 10.1038/s41467-018-03165-4} {\bibfield
   {journal} {\bibinfo  {journal} {Nature Communications}\ }\textbf {\bibinfo
  {volume} {9}},\ \bibinfo {eid} {795} (\bibinfo {year} {2018})},\ \Eprint
  {http://arxiv.org/abs/1704.01080} {arXiv:1704.01080 [hep-ex]} \BibitemShut
  {NoStop}%
\bibitem [{\citenamefont {{Lynch}}(2024)}]{2024PhRvD.109j5009L}%
  \BibitemOpen
  \bibfield  {author} {\bibinfo {author} {\bibfnamefont {Morgan~H.}\
  \bibnamefont {{Lynch}}},\ }\bibfield  {title} {\enquote {\bibinfo {title}
  {{Analysis of the CERN-NA63 radiation reaction data set, assuming the Rindler
  bath is composed of microscopic black holes}},}\ }\href {\doibase
  10.1103/PhysRevD.109.105009} {\bibfield  {journal} {\bibinfo  {journal}
  {\prd}\ }\textbf {\bibinfo {volume} {109}},\ \bibinfo {eid} {105009}
  (\bibinfo {year} {2024})},\ \Eprint {http://arxiv.org/abs/2404.09274}
  {arXiv:2404.09274 [gr-qc]} \BibitemShut {NoStop}%
\bibitem [{\citenamefont {{Wistisen}}\ \emph {et~al.}(2019)\citenamefont
  {{Wistisen}}, \citenamefont {{Di Piazza}}, \citenamefont {{Nielsen}},
  \citenamefont {{S{\o}rensen}}, \citenamefont {{Uggerh{\o}j}},\ and\
  \citenamefont {{CERN NA63 Collaboration}}}]{2019PhRvR...1c3014W}%
  \BibitemOpen
  \bibfield  {author} {\bibinfo {author} {\bibfnamefont {T.~N.}\ \bibnamefont
  {{Wistisen}}}, \bibinfo {author} {\bibfnamefont {A.}~\bibnamefont {{Di
  Piazza}}}, \bibinfo {author} {\bibfnamefont {C.~F.}\ \bibnamefont
  {{Nielsen}}}, \bibinfo {author} {\bibfnamefont {A.~H.}\ \bibnamefont
  {{S{\o}rensen}}}, \bibinfo {author} {\bibfnamefont {U.~I.}\ \bibnamefont
  {{Uggerh{\o}j}}}, \ and\ \bibinfo {author} {\bibnamefont {{CERN NA63
  Collaboration}}},\ }\bibfield  {title} {\enquote {\bibinfo {title} {{Quantum
  radiation reaction in aligned crystals beyond the local constant field
  approximation}},}\ }\href {\doibase 10.1103/PhysRevResearch.1.033014}
  {\bibfield  {journal} {\bibinfo  {journal} {Physical Review Research}\
  }\textbf {\bibinfo {volume} {1}},\ \bibinfo {eid} {033014} (\bibinfo {year}
  {2019})},\ \Eprint {http://arxiv.org/abs/1906.09144} {arXiv:1906.09144
  [physics.plasm-ph]} \BibitemShut {NoStop}%
\bibitem [{\citenamefont {{Di Piazza}}\ \emph {et~al.}(2012)\citenamefont {{Di
  Piazza}}, \citenamefont {{M{\"u}ller}}, \citenamefont {{Hatsagortsyan}},\
  and\ \citenamefont {{Keitel}}}]{2012RvMP...84.1177D}%
  \BibitemOpen
  \bibfield  {author} {\bibinfo {author} {\bibfnamefont {A.}~\bibnamefont {{Di
  Piazza}}}, \bibinfo {author} {\bibfnamefont {C.}~\bibnamefont
  {{M{\"u}ller}}}, \bibinfo {author} {\bibfnamefont {K.~Z.}\ \bibnamefont
  {{Hatsagortsyan}}}, \ and\ \bibinfo {author} {\bibfnamefont {C.~H.}\
  \bibnamefont {{Keitel}}},\ }\bibfield  {title} {\enquote {\bibinfo {title}
  {{Extremely high-intensity laser interactions with fundamental quantum
  systems}},}\ }\href {\doibase 10.1103/RevModPhys.84.1177} {\bibfield
  {journal} {\bibinfo  {journal} {Reviews of Modern Physics}\ }\textbf
  {\bibinfo {volume} {84}},\ \bibinfo {pages} {1177--1228} (\bibinfo {year}
  {2012})},\ \Eprint {http://arxiv.org/abs/1111.3886} {arXiv:1111.3886
  [hep-ph]} \BibitemShut {NoStop}%
\bibitem [{\citenamefont {{Cozzella}}\ \emph {et~al.}(2017)\citenamefont
  {{Cozzella}}, \citenamefont {{Landulfo}}, \citenamefont {{Matsas}},\ and\
  \citenamefont {{Vanzella}}}]{2017PhRvL.118p1102C}%
  \BibitemOpen
  \bibfield  {author} {\bibinfo {author} {\bibfnamefont {Gabriel}\ \bibnamefont
  {{Cozzella}}}, \bibinfo {author} {\bibfnamefont {Andr{\'e} G.~S.}\
  \bibnamefont {{Landulfo}}}, \bibinfo {author} {\bibfnamefont {George E.~A.}\
  \bibnamefont {{Matsas}}}, \ and\ \bibinfo {author} {\bibfnamefont {Daniel
  A.~T.}\ \bibnamefont {{Vanzella}}},\ }\bibfield  {title} {\enquote {\bibinfo
  {title} {{Proposal for Observing the Unruh Effect using Classical
  Electrodynamics}},}\ }\href {\doibase 10.1103/PhysRevLett.118.161102}
  {\bibfield  {journal} {\bibinfo  {journal} {\prl}\ }\textbf {\bibinfo
  {volume} {118}},\ \bibinfo {eid} {161102} (\bibinfo {year} {2017})},\ \Eprint
  {http://arxiv.org/abs/1701.03446} {arXiv:1701.03446 [gr-qc]} \BibitemShut
  {NoStop}%
\bibitem [{\citenamefont {{Lynch}}(2025)}]{2025arXiv250521292L}%
  \BibitemOpen
  \bibfield  {author} {\bibinfo {author} {\bibfnamefont {Morgan~H.}\
  \bibnamefont {{Lynch}}},\ }\bibfield  {title} {\enquote {\bibinfo {title}
  {{Hyperbolic recoil and the Unruh effect at CERN-NA63}},}\ }\href {\doibase
  10.48550/arXiv.2505.21292} {\bibfield  {journal} {\bibinfo  {journal} {arXiv
  e-prints}\ ,\ \bibinfo {eid} {arXiv:2505.21292}} (\bibinfo {year} {2025})},\
  \Eprint {http://arxiv.org/abs/2505.21292} {arXiv:2505.21292 [hep-ph]}
  \BibitemShut {NoStop}%
\bibitem [{\citenamefont {{Abbasov}}\ \emph {et~al.}(1986)\citenamefont
  {{Abbasov}}, \citenamefont {{Bolotovski{\u{i}}}},\ and\ \citenamefont
  {{Davydov}}}]{1986SvPhU..29..788A}%
  \BibitemOpen
  \bibfield  {author} {\bibinfo {author} {\bibfnamefont {I.~I.}\ \bibnamefont
  {{Abbasov}}}, \bibinfo {author} {\bibfnamefont {Boris~M.}\ \bibnamefont
  {{Bolotovski{\u{i}}}}}, \ and\ \bibinfo {author} {\bibfnamefont {V.~A.}\
  \bibnamefont {{Davydov}}},\ }\bibfield  {title} {\enquote {\bibinfo {title}
  {{FROM THE HISTORY OF PHYSICS: High-frequency asymptotic behavior of
  radiation spectra of moving charges in classical electrodynamics}},}\ }\href
  {\doibase 10.1070/PU1986v029n08ABEH003484} {\bibfield  {journal} {\bibinfo
  {journal} {Soviet Physics Uspekhi}\ }\textbf {\bibinfo {volume} {29}},\
  \bibinfo {pages} {788--796} (\bibinfo {year} {1986})}\BibitemShut {NoStop}%
\bibitem [{\citenamefont {{Bianchi}}\ and\ \citenamefont
  {{Satz}}(2013)}]{2013PhRvD..87l4031B}%
  \BibitemOpen
  \bibfield  {author} {\bibinfo {author} {\bibfnamefont {Eugenio}\ \bibnamefont
  {{Bianchi}}}\ and\ \bibinfo {author} {\bibfnamefont {Alejandro}\ \bibnamefont
  {{Satz}}},\ }\bibfield  {title} {\enquote {\bibinfo {title} {{Mechanical laws
  of the Rindler horizon}},}\ }\href {\doibase 10.1103/PhysRevD.87.124031}
  {\bibfield  {journal} {\bibinfo  {journal} {\prd}\ }\textbf {\bibinfo
  {volume} {87}},\ \bibinfo {eid} {124031} (\bibinfo {year} {2013})},\ \Eprint
  {http://arxiv.org/abs/1305.4986} {arXiv:1305.4986 [gr-qc]} \BibitemShut
  {NoStop}%
\bibitem [{\citenamefont {{Paithankar}}\ and\ \citenamefont
  {{Kolekar}}(2020)}]{2020PhRvD.101f5012P}%
  \BibitemOpen
  \bibfield  {author} {\bibinfo {author} {\bibfnamefont {Kajol}\ \bibnamefont
  {{Paithankar}}}\ and\ \bibinfo {author} {\bibfnamefont {Sanved}\ \bibnamefont
  {{Kolekar}}},\ }\bibfield  {title} {\enquote {\bibinfo {title} {{Role of the
  Unruh effect in Bremsstrahlung}},}\ }\href {\doibase
  10.1103/PhysRevD.101.065012} {\bibfield  {journal} {\bibinfo  {journal}
  {\prd}\ }\textbf {\bibinfo {volume} {101}},\ \bibinfo {eid} {065012}
  (\bibinfo {year} {2020})},\ \Eprint {http://arxiv.org/abs/2001.03078}
  {arXiv:2001.03078 [gr-qc]} \BibitemShut {NoStop}%
\bibitem [{\citenamefont {{Alonso-Serrano}}\ and\ \citenamefont
  {{Visser}}(2016)}]{2016PhLB..757..383A}%
  \BibitemOpen
  \bibfield  {author} {\bibinfo {author} {\bibfnamefont {Ana}\ \bibnamefont
  {{Alonso-Serrano}}}\ and\ \bibinfo {author} {\bibfnamefont {Matt}\
  \bibnamefont {{Visser}}},\ }\bibfield  {title} {\enquote {\bibinfo {title}
  {{On burning a lump of coal}},}\ }\href {\doibase
  10.1016/j.physletb.2016.04.023} {\bibfield  {journal} {\bibinfo  {journal}
  {Physics Letters B}\ }\textbf {\bibinfo {volume} {757}},\ \bibinfo {pages}
  {383--386} (\bibinfo {year} {2016})},\ \Eprint
  {http://arxiv.org/abs/1511.01162} {arXiv:1511.01162 [gr-qc]} \BibitemShut
  {NoStop}%
\bibitem [{\citenamefont {{Alonso-Serrano}}\ and\ \citenamefont
  {{Visser}}(2018)}]{2018PhLB..776...10A}%
  \BibitemOpen
  \bibfield  {author} {\bibinfo {author} {\bibfnamefont {Ana}\ \bibnamefont
  {{Alonso-Serrano}}}\ and\ \bibinfo {author} {\bibfnamefont {Matt}\
  \bibnamefont {{Visser}}},\ }\bibfield  {title} {\enquote {\bibinfo {title}
  {{Entropy/information flux in Hawking radiation}},}\ }\href {\doibase
  10.1016/j.physletb.2017.11.020} {\bibfield  {journal} {\bibinfo  {journal}
  {Physics Letters B}\ }\textbf {\bibinfo {volume} {776}},\ \bibinfo {pages}
  {10--16} (\bibinfo {year} {2018})}\BibitemShut {NoStop}%
\bibitem [{\citenamefont {{Einstein}}(1907)}]{einstein..}%
  \BibitemOpen
  \bibfield  {author} {\bibinfo {author} {\bibfnamefont {A.}~\bibnamefont
  {{Einstein}}},\ }\bibfield  {title} {\enquote {\bibinfo {title} {{Jahrbuch d.
  Radioaktivitat und Elektronik}},}\ }\href@noop {} {\  (\bibinfo {year}
  {1907})}\BibitemShut {NoStop}%
\bibitem [{\citenamefont {{Landsberg}}(1966)}]{1966Natur.212..571L}%
  \BibitemOpen
  \bibfield  {author} {\bibinfo {author} {\bibfnamefont {P.~T.}\ \bibnamefont
  {{Landsberg}}},\ }\bibfield  {title} {\enquote {\bibinfo {title} {{Does a
  Moving Body Appear Cool?}}}\ }\href {\doibase 10.1038/212571a0} {\bibfield
  {journal} {\bibinfo  {journal} {\nat}\ }\textbf {\bibinfo {volume} {212}},\
  \bibinfo {pages} {571--572} (\bibinfo {year} {1966})}\BibitemShut {NoStop}%
\bibitem [{\citenamefont {{Landsberg}}(1967)}]{1967Natur.214..903L}%
  \BibitemOpen
  \bibfield  {author} {\bibinfo {author} {\bibfnamefont {P.~T.}\ \bibnamefont
  {{Landsberg}}},\ }\bibfield  {title} {\enquote {\bibinfo {title} {{Does a
  Moving Body appear Cool?}}}\ }\href {\doibase 10.1038/214903a0} {\bibfield
  {journal} {\bibinfo  {journal} {\nat}\ }\textbf {\bibinfo {volume} {214}},\
  \bibinfo {pages} {903--904} (\bibinfo {year} {1967})}\BibitemShut {NoStop}%
\bibitem [{\citenamefont {{Far{\'\i}as}}\ \emph {et~al.}(2017)\citenamefont
  {{Far{\'\i}as}}, \citenamefont {{Pinto}},\ and\ \citenamefont
  {{Moya}}}]{2017NatSR...717657F}%
  \BibitemOpen
  \bibfield  {author} {\bibinfo {author} {\bibfnamefont {Cristian}\
  \bibnamefont {{Far{\'\i}as}}}, \bibinfo {author} {\bibfnamefont {Victor~A.}\
  \bibnamefont {{Pinto}}}, \ and\ \bibinfo {author} {\bibfnamefont {Pablo~S.}\
  \bibnamefont {{Moya}}},\ }\bibfield  {title} {\enquote {\bibinfo {title}
  {{What is the temperature of a moving body?}}}\ }\href {\doibase
  10.1038/s41598-017-17526-4} {\bibfield  {journal} {\bibinfo  {journal}
  {Scientific Reports}\ }\textbf {\bibinfo {volume} {7}},\ \bibinfo {eid}
  {17657} (\bibinfo {year} {2017})}\BibitemShut {NoStop}%
\bibitem [{\citenamefont {{Biermann}}\ \emph {et~al.}(2020)\citenamefont
  {{Biermann}}, \citenamefont {{Erne}}, \citenamefont {{Gooding}},
  \citenamefont {{Louko}}, \citenamefont {{Schmiedmayer}}, \citenamefont
  {{Unruh}},\ and\ \citenamefont {{Weinfurtner}}}]{2020PhRvD.102h5006B}%
  \BibitemOpen
  \bibfield  {author} {\bibinfo {author} {\bibfnamefont {Steffen}\ \bibnamefont
  {{Biermann}}}, \bibinfo {author} {\bibfnamefont {Sebastian}\ \bibnamefont
  {{Erne}}}, \bibinfo {author} {\bibfnamefont {Cisco}\ \bibnamefont
  {{Gooding}}}, \bibinfo {author} {\bibfnamefont {Jorma}\ \bibnamefont
  {{Louko}}}, \bibinfo {author} {\bibfnamefont {J{\"o}rg}\ \bibnamefont
  {{Schmiedmayer}}}, \bibinfo {author} {\bibfnamefont {William~G.}\
  \bibnamefont {{Unruh}}}, \ and\ \bibinfo {author} {\bibfnamefont {Silke}\
  \bibnamefont {{Weinfurtner}}},\ }\bibfield  {title} {\enquote {\bibinfo
  {title} {{Unruh and analogue Unruh temperatures for circular motion in 3 +1
  and 2 +1 dimensions}},}\ }\href {\doibase 10.1103/PhysRevD.102.085006}
  {\bibfield  {journal} {\bibinfo  {journal} {\prd}\ }\textbf {\bibinfo
  {volume} {102}},\ \bibinfo {eid} {085006} (\bibinfo {year} {2020})},\ \Eprint
  {http://arxiv.org/abs/2007.09523} {arXiv:2007.09523 [gr-qc]} \BibitemShut
  {NoStop}%
\bibitem [{\citenamefont {{Landsberg}}\ and\ \citenamefont
  {{DeVos}}(1989)}]{1989JPhA...22.1073L}%
  \BibitemOpen
  \bibfield  {author} {\bibinfo {author} {\bibfnamefont {P.~T.}\ \bibnamefont
  {{Landsberg}}}\ and\ \bibinfo {author} {\bibfnamefont {A.}~\bibnamefont
  {{DeVos}}},\ }\bibfield  {title} {\enquote {\bibinfo {title} {{The
  Stefan-Boltzmann constant in n-dimensional space}},}\ }\href {\doibase
  10.1088/0305-4470/22/8/021} {\bibfield  {journal} {\bibinfo  {journal}
  {Journal of Physics A Mathematical General}\ }\textbf {\bibinfo {volume}
  {22}},\ \bibinfo {pages} {1073--1084} (\bibinfo {year} {1989})}\BibitemShut
  {NoStop}%
\bibitem [{\citenamefont {Bekenstein}\ and\ \citenamefont
  {Mayo}(2001)}]{Bekenstein:2001tj}%
  \BibitemOpen
  \bibfield  {author} {\bibinfo {author} {\bibfnamefont {Jacob~D.}\
  \bibnamefont {Bekenstein}}\ and\ \bibinfo {author} {\bibfnamefont
  {Avraham~E.}\ \bibnamefont {Mayo}},\ }\bibfield  {title} {\enquote {\bibinfo
  {title} {{Black holes are one-dimensional}},}\ }\href {\doibase
  10.1023/A:1015278813573} {\bibfield  {journal} {\bibinfo  {journal} {Gen.
  Rel. Grav.}\ }\textbf {\bibinfo {volume} {33}},\ \bibinfo {pages}
  {2095--2099} (\bibinfo {year} {2001})},\ \Eprint
  {http://arxiv.org/abs/gr-qc/0105055} {arXiv:gr-qc/0105055} \BibitemShut
  {NoStop}%
\bibitem [{\citenamefont {{Nielsen}}\ \emph {et~al.}(2023)\citenamefont
  {{Nielsen}}, \citenamefont {{Holtzapple}}, \citenamefont {{Lund}},
  \citenamefont {{Surrow}}, \citenamefont {{S{\o}rensen}}, \citenamefont
  {{S{\o}rensen}}, \citenamefont {{Uggerh{\o}j}},\ and\ \citenamefont {{CERN
  NA63}}}]{2023PhRvL.130g1601N}%
  \BibitemOpen
  \bibfield  {author} {\bibinfo {author} {\bibfnamefont {Christian~F.}\
  \bibnamefont {{Nielsen}}}, \bibinfo {author} {\bibfnamefont {Robert}\
  \bibnamefont {{Holtzapple}}}, \bibinfo {author} {\bibfnamefont {Mads~M.}\
  \bibnamefont {{Lund}}}, \bibinfo {author} {\bibfnamefont {Jeppe~H.}\
  \bibnamefont {{Surrow}}}, \bibinfo {author} {\bibfnamefont {Allan~H.}\
  \bibnamefont {{S{\o}rensen}}}, \bibinfo {author} {\bibfnamefont {Marc~B.}\
  \bibnamefont {{S{\o}rensen}}}, \bibinfo {author} {\bibfnamefont {Ulrik~I.}\
  \bibnamefont {{Uggerh{\o}j}}}, \ and\ \bibinfo {author} {\bibnamefont {{CERN
  NA63}}},\ }\bibfield  {title} {\enquote {\bibinfo {title} {{Precision
  Measurement of Trident Production in Strong Electromagnetic Fields}},}\
  }\href {\doibase 10.1103/PhysRevLett.130.071601} {\bibfield  {journal}
  {\bibinfo  {journal} {\prl}\ }\textbf {\bibinfo {volume} {130}},\ \bibinfo
  {eid} {071601} (\bibinfo {year} {2023})},\ \Eprint
  {http://arxiv.org/abs/2211.02390} {arXiv:2211.02390 [hep-ex]} \BibitemShut
  {NoStop}%
\bibitem [{\citenamefont {Workman}\ \emph {et~al.}(2022)\citenamefont {Workman}
  \emph {et~al.}}]{Workman:2022ynf}%
  \BibitemOpen
  \bibfield  {author} {\bibinfo {author} {\bibfnamefont {R.~L.}\ \bibnamefont
  {Workman}} \emph {et~al.} (\bibinfo {collaboration} {Particle Data Group}),\
  }\bibfield  {title} {\enquote {\bibinfo {title} {{Review of Particle
  Physics}},}\ }\href {\doibase 10.1093/ptep/ptac097} {\bibfield  {journal}
  {\bibinfo  {journal} {PTEP}\ }\textbf {\bibinfo {volume} {2022}},\ \bibinfo
  {pages} {083C01} (\bibinfo {year} {2022})}\BibitemShut {NoStop}%
\end{thebibliography}%

\end{document}